\documentclass[aps,prb,twocolumn,superscriptaddress,floatfix,longbibliography]{revtex4-2}

\usepackage{amsmath,amssymb} % math symbols
\usepackage{bm} % bold math font
\usepackage{graphicx} % for figures
\usepackage{comment} % allows block comments
\usepackage{textcomp} % This package is just to give the text quote '
\usepackage[nolist,nohyperlinks]{acronym}
\usepackage{float}
\usepackage{upgreek}
\usepackage{siunitx}
\usepackage{enumitem}
\setlist{noitemsep,leftmargin=*,topsep=0pt,parsep=0pt}

\usepackage{xcolor} % \textcolor{red}{text} will be red for notes
\definecolor{lightgray}{gray}{0.6}
\definecolor{medgray}{gray}{0.4}

\usepackage{hyperref}
\hypersetup{
colorlinks=true,
urlcolor= blue,
citecolor=blue,
linkcolor= blue,
}

\newif\ifptitle
\newif\ifpnumber
\newcounter{para}

\ptitletrue  % comment this line to hide paragraph titles
\pnumbertrue  % comment this line to hide paragraph numbers

\newcommand{\mytitle}{Nozaki-Bekki optical solitons}

\begin{document}

\title{\mytitle}

\author{Nikola Opa$\mathrm{\check{c}}$ak}
\email[]{nikola.opacak@tuwien.ac.at}
\affiliation{Institute of Solid State Electronics, TU Wien, 1040 Vienna, Austria}
\affiliation{Harvard John A. Paulson School of Engineering and Applied Sciences, Harvard University, Cambridge, MA 02138, USA}

\author{Dmitry Kazakov}
%\email[]{kazakov@seas.harvard.edu}
\affiliation{Harvard John A. Paulson School of Engineering and Applied Sciences, Harvard University, Cambridge, MA 02138, USA}

\author{Lorenzo L. Columbo}
\affiliation{Dipartimento di Elettronica e Telecomunicazioni, Politecnico di Torino, 10129 Torino, Italy}

\author{Maximilian Beiser}
\affiliation{Institute of Solid State Electronics, TU Wien, 1040 Vienna, Austria}

\author{Theodore P. Letsou}
\affiliation{Harvard John A. Paulson School of Engineering and Applied Sciences, Harvard University, Cambridge, MA 02138, USA}%\usepackage[nolist,nohyperlinks]{acronym}
 \affiliation{Department of Electrical Engineering and Computer Science, Massachusetts Institute of Technology, Cambridge, MA 02142, USA}

\author{Florian Pilat}
\affiliation{Institute of Solid State Electronics, TU Wien, 1040 Vienna, Austria}

% \author{Yongrui Wang}
% \affiliation{Department of Physics and Astronomy, Texas A\&M University, College Station, TX 77843, USA}

% \author{Alexey Belyanin}
% \affiliation{Department of Physics and Astronomy, Texas A\&M University, College Station, TX 77843, USA}

\author{Massimo Brambilla}
\affiliation{Dipartimento di Fisica Interateneo and CNR-IFN, Università e Politecnico di Bari, 70125 Bari, Italy}

\author{Franco Prati}
\affiliation{Dipartimento di Scienza e Alta Tecnologia, Università dell’Insubria, 22100 Como, Italy}

\author{Marco Piccardo}
% \email[]{piccardo@g.harvard.edu}
\affiliation{Harvard John A. Paulson School of Engineering and Applied Sciences, Harvard University, Cambridge, MA 02138, USA}
\affiliation{Department of Physics, Instituto Superior Técnico, Universidade de Lisboa, 1049-001 Lisbon, Portugal}
\affiliation{Instituto de Engenharia de Sistemas e Computadores -- Microsistemas e Nanotecnologias (INESC MN), 1000-029 Lisbon, Portugal}

\author{Federico Capasso}
%\email[]{capasso@seas.harvard.edu}
\affiliation{Harvard John A. Paulson School of Engineering and Applied Sciences, Harvard University, Cambridge, MA 02138, USA}

\author{Benedikt Schwarz}
% \email[]{benedikt.schwarz@tuwien.ac.at}
\affiliation{Institute of Solid State Electronics, TU Wien, 1040 Vienna, Austria}
\affiliation{Harvard John A. Paulson School of Engineering and Applied Sciences, Harvard University, Cambridge, MA 02138, USA}

%\date{\today}

\begin{abstract}

Recent years witnessed rapid progress of chip-scale integrated optical frequency comb sources.
Among them, two classes are particularly significant
-- semiconductor Fabry-Per\'{o}t lasers and passive ring Kerr microresonators. 
Here, we merge the two technologies in a ring semiconductor laser and demonstrate a new paradigm for free-running soliton formation, called Nozaki-Bekki soliton. 
These dissipative waveforms emerge in a family of traveling localized dark pulses, known within the famed complex Ginzburg-Landau equation.
We show that Nozaki-Bekki solitons are structurally-stable in a ring laser and form spontaneously with tuning of the laser bias -- eliminating the need for an external optical pump. 
By combining conclusive experimental findings and a complementary elaborate theoretical model, we reveal the salient characteristics of these solitons and provide a guideline for their generation. 
Beyond the fundamental soliton circulating inside the ring laser, 
we demonstrate multisoliton states as well, verifying their localized nature and offering an insight into formation of soliton crystals. 
Our results consolidate a monolithic electrically-driven platform for direct soliton generation and open a door for a new research field at the junction of laser multimode dynamics and Kerr parametric processes.

\end{abstract}

\maketitle

%Stable, self-organizing, and 
Dissipative temporal solitons -- stable solitary localized pulses -- emerge universally 
in extended nonlinear media, sustained by a dual balance between the nonlinearity and dispersion/diffusion, as well as between the gain and dissipation of the system~\cite{akhmediev2008dissipative}.
Their prime examples in optics came from passively mode-locked lasers~\cite{grelu2012dissipative}, optical fibers~\cite{englebert2021temporal,leo2010temporal}, and passive high-\textit{Q} microresonators~\cite{herr2013temporal,kippenberg2018dissipative,rowley2022self}. 
Being of special interest for integrated photonics due to their compact size, microresonators have taken the community by storm.  
Fine-tuning of the requisite \ac{cw} optical pump, combined with a bulk Kerr nonlinearity, is necessary to provide the parametric gain that leads to bright or dark microresonator solitons~\cite{herr2012universal,guo2016universal,xue2015mode,zhang2022dark}. 
When outcoupled, the circulating soliton results in 
a periodic train of short pulses, generating a broad frequency comb in the spectral domain~\cite{udem2002optical}. These miniature Kerr combs have ever since been 
the vanguard of microresonator technology, finding use in telecommunication~\cite{marin_palomo2017microresonator,fulop2018highorder}, ultrafast ranging~\cite{trocha2018ultrafast}, high-precision spectroscopy~\cite{picque2019frequency,suh2016microresonator}, and frequency synthesis~\cite{spencer2018optical}.

%New section
In this work, we demonstrate a new type of optical dissipative solitons -- named \ac{nb} solitons -- in an electrically-driven \ac{mir} ring semiconductor laser. 
They arise as localized waveforms in a family of traveling dark pulses that satisfy ring periodic boundary conditions. 
The active laser gain material in our devices simultaneously provides a giant Kerr nonlinearity and eliminates the external optical pump and its challenging frequency tuning, which is a vital ingredient for microresonator combs.
The \ac{nb} soliton regime -- first of its kind in a compact optical system -- emerges spontaneously and is directly accessed solely by tuning the laser driving current, which we validate by using a phase-sensitive measurement.
The experimental findings are corroborated by a complementary theoretical Maxwell-Bloch formalism with numerical simulations, allowing us to identify favourable dispersive and nonlinear conditions for \ac{nb} soliton formation and their coherent control. 
Our initial prediction of their existence originated from the cubic \ac{cgle} 
-- one of the most celebrated equations in physics that describes spatially extended systems close to bifurcations~\cite{aranson2002world}. While their stability was long discussed within the \ac{cgle} framework, \ac{nb} solitons have so far been scarcely observed in experiments. 
Unequivocal classification of \ac{nb} solitons is made possible by their striking salient characteristics -- anti-phase synchronization of the soliton with the primary mode and a $2\pi$ temporal phase ramp across the soliton. 
The localized nature of \ac{nb} solitons is conclusively demonstrated by further observing multisoliton states, both in theory and experiments.
Our findings herald a new generation of monolithically-integrated self-starting soliton generators that lie at the intersection of semiconductor lasers and Kerr microresonators.

\begin{figure}[t!]
	\centering
	\includegraphics[width = 1\columnwidth]{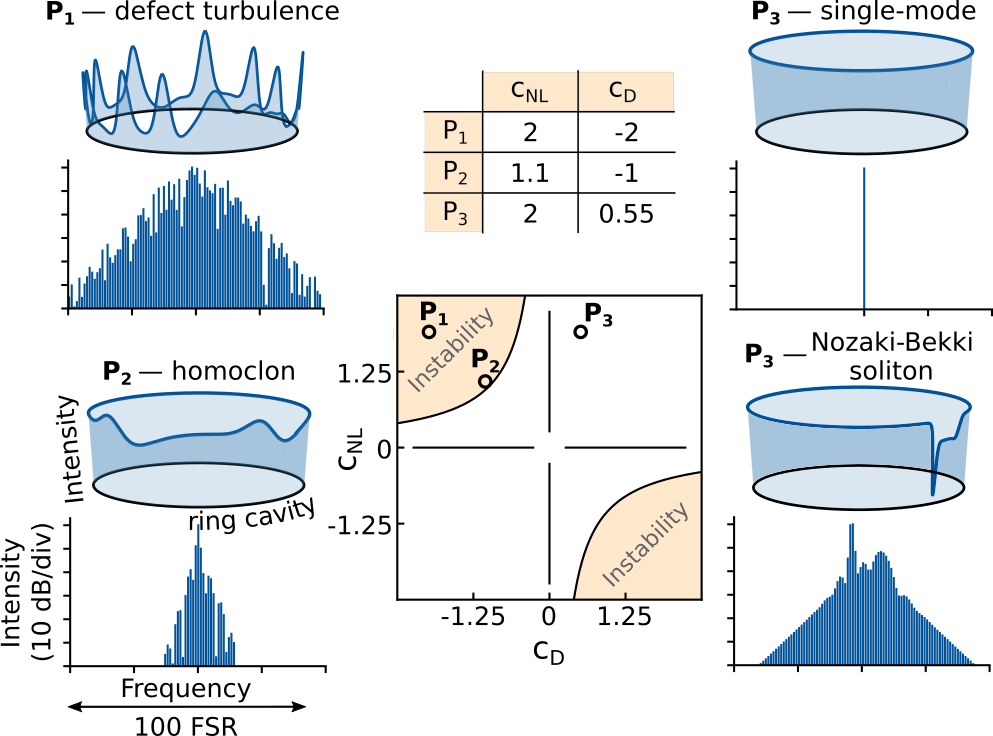}
	\caption{ \textbf{Parameter space of the CGLE with corresponding laser regimes.} The unidirectional intracavity intensities and the spectra are obtained from numerical simulations of the CGLE for different points indicated in the parameter space. The left column shows two states triggered by turbulence inside the CW linearly-unstable yellow region of the parameter space (defined by $1+c_{\mathrm{D}}c_{\mathrm{NL}}<0$), where a single-mode field cannot exist as a steady state. Defect turbulence occurs deep inside the unstable region (point $P_1$) and is aperiodic, exhibiting chaotic temporal evolution. Close to the border of CW stability (point $P_2$), phase turbulence leads to narrowband homoclon frequency combs~\cite{piccardo2020frequency}. The right column displays the coexistence (multistability) of a single-mode regime and an NB soliton, both set in the CW linearly-stable part of the parameter space ($1+c_{\mathrm{D}}c_{\mathrm{NL}}>0$). NB solitons emerge as coherent, unidirectional propagating dark pulses, characterized with a broad and smooth spectral envelope. The visible pulse shoulder represents a shock, stabilizing the waveform in ring periodic boundary conditions.}
	\label{fig1}
\end{figure}

To study \ac{nb} solitons, we use \acp{qcl} -- compact and efficient devices that emit in the \ac{mir} and THz regions~\cite{faist1994quantum,yao2012mid,williams2007thz} -- embedded in a ring cavity. 
Ultrafast intersubband transitions of \acp{qcl} provide not only the optical gain, but also a giant Kerr nonlinearity~\cite{opacak2021frequency,friedli2013four}, which is several orders of magnitude larger than in bulk III-V compounds~\cite{gaeta2019photonic}.
This is already exploited for %electrically-driven  
frequency comb formation in \ac{fp} \acp{qcl}~\cite{hugi2012mid,opacak2021frequency,opacak2019theory}. 
There, the multimode operation originates from the \ac{shb}~\cite{opacak2019theory}, which describes inhomogeneous gain saturation due to a standing wave inside the cavity, caused by the counter-propagating components of the electric field. 
Conversely, a ring cavity has no reflection points and supports unidirectional field propagation -- preventing \ac{shb}. 
Nevertheless, recent work demonstrated multimode unidirectional ring \acp{qcl} due to phase turbulence at low pumping levels, making \ac{shb} nonessential~\cite{piccardo2020frequency}. 
Since then, substantial efforts were put towards developing Kerr combs in ring \acp{qcl}~\cite{meng2020mid,jaildl2021comb,jaidl2022silicon,micheleti2022thz}, leading even to bright pulses after spectral filtering~\cite{meng2021dissipative} and theoretically-predicted optically-driven solitons~\cite{columbo2021unifying,prati2020soliton,prati2021global} -- already anticipating the potential of these devices as an on-chip soliton platform. % for soliton generation.

%By revisiting the theory of semiconductor lasers with a fast gain medium, 
We can interpret ring QCL multimode dynamics on the grounds of the \ac{cgle} with periodic boundary conditions~\cite{piccardo2020frequency}.   
Starting from the more general laser master equation~\cite{opacak2019theory,opacak2022thesis}, 
we derive the cubic \ac{cgle} 
\begin{align}
\partial_tE = E + (1+ic_{\mathrm{D}})\partial^2_z E - (1+ic_{\mathrm{NL}})|E|^2E,\label{eq:1}
\end{align}
where $E$ is the unidirectional electric field, $t$ is time, and $z$ is the spatial coordinate along the ring cavity (see derivation in the Supplementary material). 
The entire parameter space is elegantly constricted to just two dimensions, 
which refer to the dispersive ($c_{\mathrm{D}}$) and nonlinear effects ($c_{\mathrm{NL}}$). In lasers, $c_{\mathrm{D}}$ is partly determined by the cavity \ac{gvd} $k^{''}$. Another contribution to the total \ac{gvd} originates from the gain lineshape, defined with the \ac{lef} -- a crucial semiconductor laser parameter which describes light amplitude-phase coupling~\cite{henry1982theory,opacak2021spectrally}. Moreover, the \ac{lef} also defines $c_{\mathrm{NL}}$ and phenomenologically describes the giant resonant Kerr nonlinearity of \acp{qcl}~\cite{opacak2019theory,opacak2021frequency}. %that originates from the gain itself
A linear stability analysis of the \ac{cgle} divides the $c_{\mathrm{D}}-c_{\mathrm{NL}}$ space %spun by $c_{\mathrm{D}}$ and $c_{\mathrm{NL}}$ 
in two regions depending on the stability of the single-mode \ac{cw} solution under small perturbations (Fig.~\ref{fig1})~\cite{aranson2002world}. %, given by the condition $1+c_{\mathrm{D}}c_{\mathrm{NL}}>0$.
%by the condition $1+c_{\mathrm{D}}c_{\mathrm{NL}}=0$, depending if a single-mode solution under small perturbations is stable or not (Fig.~\ref{fig1})~\cite{aranson2002world}. 
Deep within the unstable region, defect turbulence occurs, characterized with a broad, unlocked spectrum and chaotically-evolving intracavity intensity (point $\mathrm{P}_1$). 
Closer to the stable region, the laser undergoes phase turbulence represented by a narrower spectrum and shallow intensity variations (point $\mathrm{P}_2$). Phase turbulence can eventually lead to  
frequency combs in the form of localized coherent pulse-like structures named homoclons~\cite{aranson2002world,piccardo2020frequency}. 

\begin{figure*}[t]
	\centering
	\includegraphics[width = 1\textwidth]{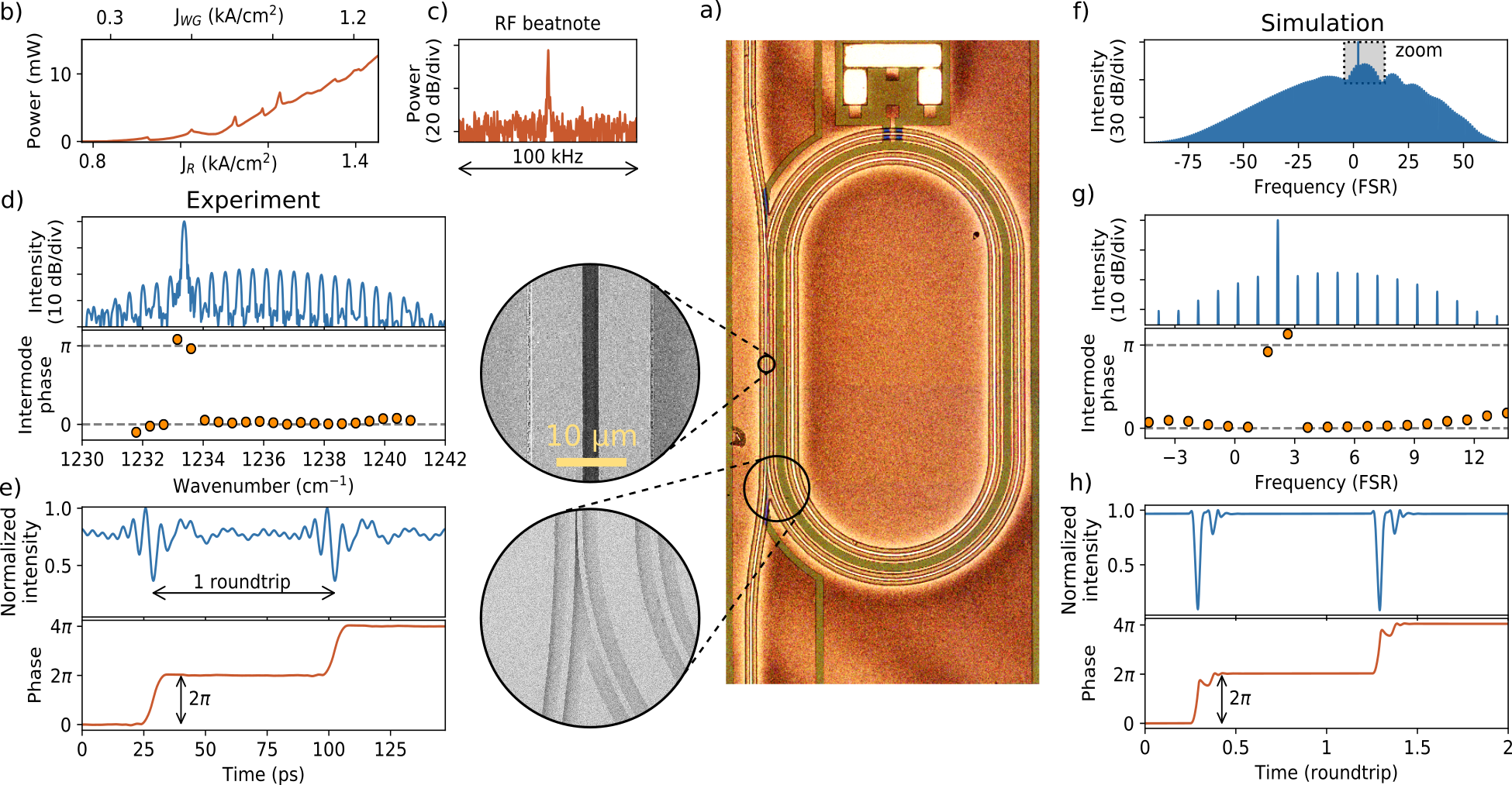}
	\caption{ \textbf{Experimental and theoretical characterization of fundamental NB solitons in a monolithic ring laser.} \textbf{a)} Microscope image of the \ac{qcl} ring and waveguide coupler, with separate electrical contacts. %The straight sections of the ring cavity are 1.5$\,\mathrm{mm}$ long, while the bend radius of the curved sections is 500$\,\mathrm{\upmu m}$. 
    SEM images depict the ring-waveguide coupling region. \textbf{b)} Output power of the device as a function of the ring and waveguide currents, $J_{\mathrm{R}}$ and $J_{\mathrm{WG}}$ respectively. %The latter is kept under the lasing threshold.
    \textbf{c)} A narrow optical beatnote of the laser comb at the central frequency of 13.59$\,\mathrm{GHz}$, equal to the repetition frequency. \textbf{d)} Experimentally-obtained intensity spectrum and the intermode phases of an \ac{nb} soliton at bias currents of $J_{\mathrm{R}}=$1.33$\,\mathrm{kA/cm}^2$ and $J_{\mathrm{WG}}=$0.79$\,\mathrm{kA/cm}^2$. Intermode phases between weaker sidemodes are synchronized in-phase, while the phase of the primary mode is $\pi$-shifted. Detailed SWIFTS measured data is presented in the Supplementary material. \textbf{e)} Temporal profiles of the intensity and phase of the emitted light. Within the width of the \ac{nb} soliton, the phase changes its value by $2\pi$ and remains linear during the remainder of the roundtrip, confirming that the \ac{nb} soliton is surrounded by a single-frequency constant \ac{cw} field.  \textbf{f)} Intensity spectrum obtained from numerical simulations of the master equation, and \textbf{g)} zoomed-in top-portion within the same range of 35$\,\mathrm{dB}$ as the experimental spectrum in d). The $\pi$ jumps around the primary mode -- salient characteristic of \ac{nb} solitons -- are visible in the intermode phases. \textbf{h)} The simulated temporal waveforms of the intensity and the phase over two roundtrips. Larger amplitude contrast compared to e) is attributed to the limited dynamic range of detection in experiments, which results in a finite number of spectral modes used for the temporal waveform reconstruction. The simulated \ac{nb} soliton is obtained for \ac{lef} of 1.25, in agreement with typical values in \acp{qcl}~\cite{opacak2021spectrally}. The cavity dispersion $k^{''}$ was set to 800$\,\mathrm{fs}^2/\mathrm{mm}$, which together with the gain dispersion brings the total \ac{gvd} value to about -700$\,\mathrm{fs}^2/\mathrm{mm}$ (see Supplementary material for a discussion on different contributions to the effective \ac{gvd} and nonlinearity in a laser). }
	\label{fig2}
\end{figure*}

%new section
The stable region of the parameter space sustains single-mode operation. However, we show that multimode emission can exist even here if we allow for large perturbations that are beyond the scope of linear stability analysis. 
The resulting frequency combs -- known as Nozaki-Bekki holes in the \ac{cgle} framework~\cite{bekki1985formations,aranson2002world,lega2001traveling} -- have a smooth and broad spectral envelope. In the temporal domain, they correspond to a family of traveling, localized dark pulses that preserve their shape and connect two stable \ac{cw} fields, giving a constant background. 
These waveforms exist in a narrow region of the parameter space, and have been so far related to dark solitons~\cite{efremidis2000stabilization} and experimentally observed in chemical systems~\cite{perraud1993one}, fluids~\cite{burguete1999bekki}, and long-cavity ring lasers~\cite{slepneva2019convective}.
In numerical simulations imposing ring periodic boundary conditions, states comprising an arranged hole and shock pair are structurally-stable even with inclusion of higher-order nonlinear terms that are not accounted for in the cubic \ac{cgle} and that play a role in real physical systems e.g. lasers~\cite{bekki1985formations,popp1993localized,popp1995hole}.
The coexistence of a stable CW field and a hole-shock pair (point $\mathrm{P}_3$) is a prime example of the phenomenon of multistability -- stochastic dependence of the laser state on the starting conditions -- thus corroborating the solitonic nature of these dissipative localized structures~\cite{akhmediev2008dissipative,kippenberg2018dissipative}, referred to hereafter as NB solitons in our lasers.

A microscope image of the ring \ac{qcl} %emitting around 8.2$\,\mathrm{\upmu m}$ 
is seen in Fig.~\ref{fig2}a), along with an integrated waveguide coupler. 
The waveguide core is made from the same \ac{qcl} material 
and has separate contacts for independent electrical driving. This allows to tune the mode indices, the Q-factor, the power of the extracted light, and the coupling between the waveguide and the ring -- making this configuration an ideal testbed for a plethora of 
resonant electromagnetic phenomena~\cite{kazakov2022semiconductor}.
The light outcoupling in previous experiments relied on the minuscule ring bending losses, % due to the curved ring sidewalls,
thus limiting the extracted power to submilliwatt levels at room temperature~\cite{piccardo2020frequency}. % and prohibiting any applications. 
Using the waveguide coupler to efficiently extract the light, our devices reach power levels 
%Our device configuration ring \ac{qcl} with a coupler waveguide reaches power levels 
above 10$\,\mathrm{mW}$ (Fig.~\ref{fig2}b)), bringing them on par with \ac{fp} \acp{qcl} of a similar ridge length and width, fabricated from the same wafer~\cite{kazakov2022semiconductor}.  
While the coupling waveguide is biased below the lasing threshold, 
the ring \ac{qcl} is driven above it, where it operates in a unidirectional regime after the symmetry-breaking point~\cite{meng2021dissipative,kazakov2022semiconductor}. We find that at currents partially higher than the lasing threshold $J_{\mathrm{th}}$ (around 1.2$J_{\mathrm{th}}$), the ring laser emits a multimode field with a narrow beatnote (Fig.~\ref{fig2}c)). 
This implies a high degree of coherence, typical of a frequency comb.

\begin{figure*}[t!]
	\centering
	\includegraphics[width = 1\textwidth]{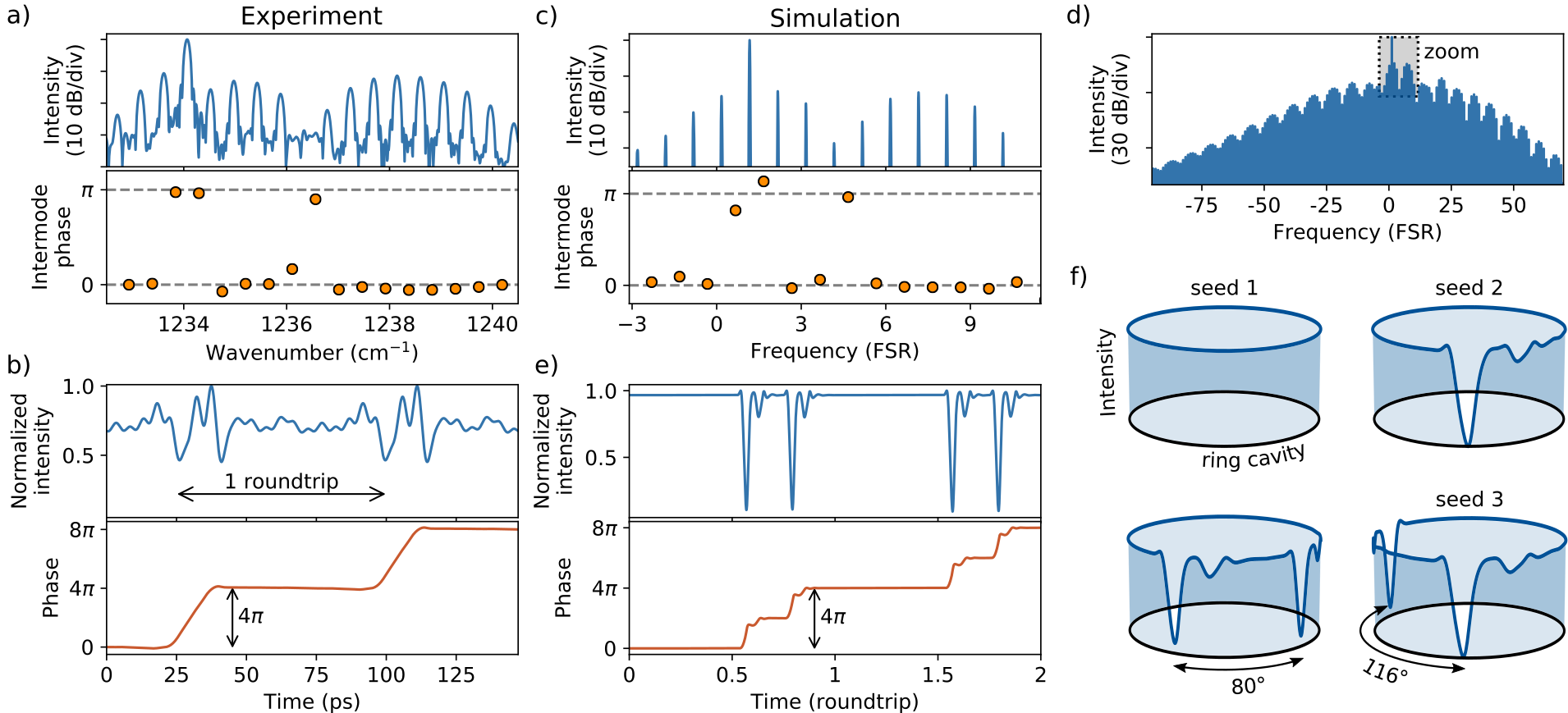}
	\caption{ \textbf{Multisoliton states.} \textbf{a)} Spectral behavior of an experimental multisoliton state, obtained for $J_{\mathrm{R}}=$1.28$\,\mathrm{kA/cm}^2$ and $J_{\mathrm{WG}}=$0.74$\,\mathrm{kA/cm}^2$. The interference between two \ac{nb} solitons causes spectral modulation, resulting in smooth lobes separated by a spectral hole. The intermode phases indicate an additional $\pi$ jump between the lobes. \textbf{b)} Two \ac{nb} solitons appear in the intensity waveform, surrounded by a \ac{cw} background. The temporal phase sweeps $4\pi$ within the two-soliton region. \textbf{c)} Zoomed top portion of the simulated spectrum and the intermode phases, in agreement with a). \textbf{d)} The intensity spectrum with the indicated zoomed section, shown in c). The spectral modulation, caused by the interference of the two solitons, cascades through the entire spectrum. \textbf{e)} Simulated temporal waveforms of the intensity and the phase. The larger dynamical range compared to the experiment allowed us to distinguish two individual $2\pi$ ramps in the phase profile, one for each soliton. The \ac{lef} was set to 1.35 and $k^{''}$ to 600$\,\mathrm{fs}^2/\mathrm{mm}$ (total \ac{gvd} around -900$\,\mathrm{fs}^2/\mathrm{mm}$). \textbf{f)} Intracavity intensity profiles obtained for the same laser parameters when starting from different random noise conditions, labeled with seed 1,2, and 3. The coexistence of a \ac{cw} single-mode field, a fundamental \ac{nb} soliton, and a multisoliton state verifies multistability and the fact that the laser operates in a linearly-stable parameter region. The distance between the \ac{nb} solitons in a multisoliton state is time-invariant, but it changes for different laser parameters or starting conditions, as is seen from two states in the bottom row. The state in the bottom left is taken from e). Experimental evidence of a soliton crystal, which is a special case of multisoliton states where all of the solitons are equidistant~\cite{karpov2019dynamics}, is found in the Supplementary material.  }
	\label{fig3}
\end{figure*}

To benchmark comb operation, we employ SWIFTS -- an experimental technique that extracts both the amplitude and phase of spectral modes~\cite{burghoff2015evaluating}. 
The measured intensity spectrum (Fig.~\ref{fig2}d)) consists of a strong primary mode surrounded by weaker sidemodes that form a smooth envelope, strikingly reminiscent of soliton spectra in microresonators~\cite{herr2013temporal,kippenberg2018dissipative,guo2016universal,xue2015mode,zhang2022dark,rowley2022self}. 
The intermode phases -- the phase differences between adjacent modes -- are shown below. They are all synchronized in-phase, except around the primary mode, where $\pi$ jumps indicate destructive interference due to anti-phase synchronization. 
This is evident from the reconstructed intensity profile (Fig.~\ref{fig2}e)), where a single dark pulse circulates around the cavity in an otherwise quasi-constant \ac{cw} background -- consistent with the predicted fundamental \ac{nb} soliton. Residual intensity oscillations are due to the limited detection bandwidth. 
Within the width of the \ac{nb} soliton, the temporal phase exhibits a steep ramp covering $2\pi$ and remains linear everywhere else -- proving that the \ac{cw} background around the soliton is constant and contains a single optical frequency equal to that of the primary mode. 

\begin{figure*}[t!]
	\centering
	\includegraphics[width = 1\textwidth]{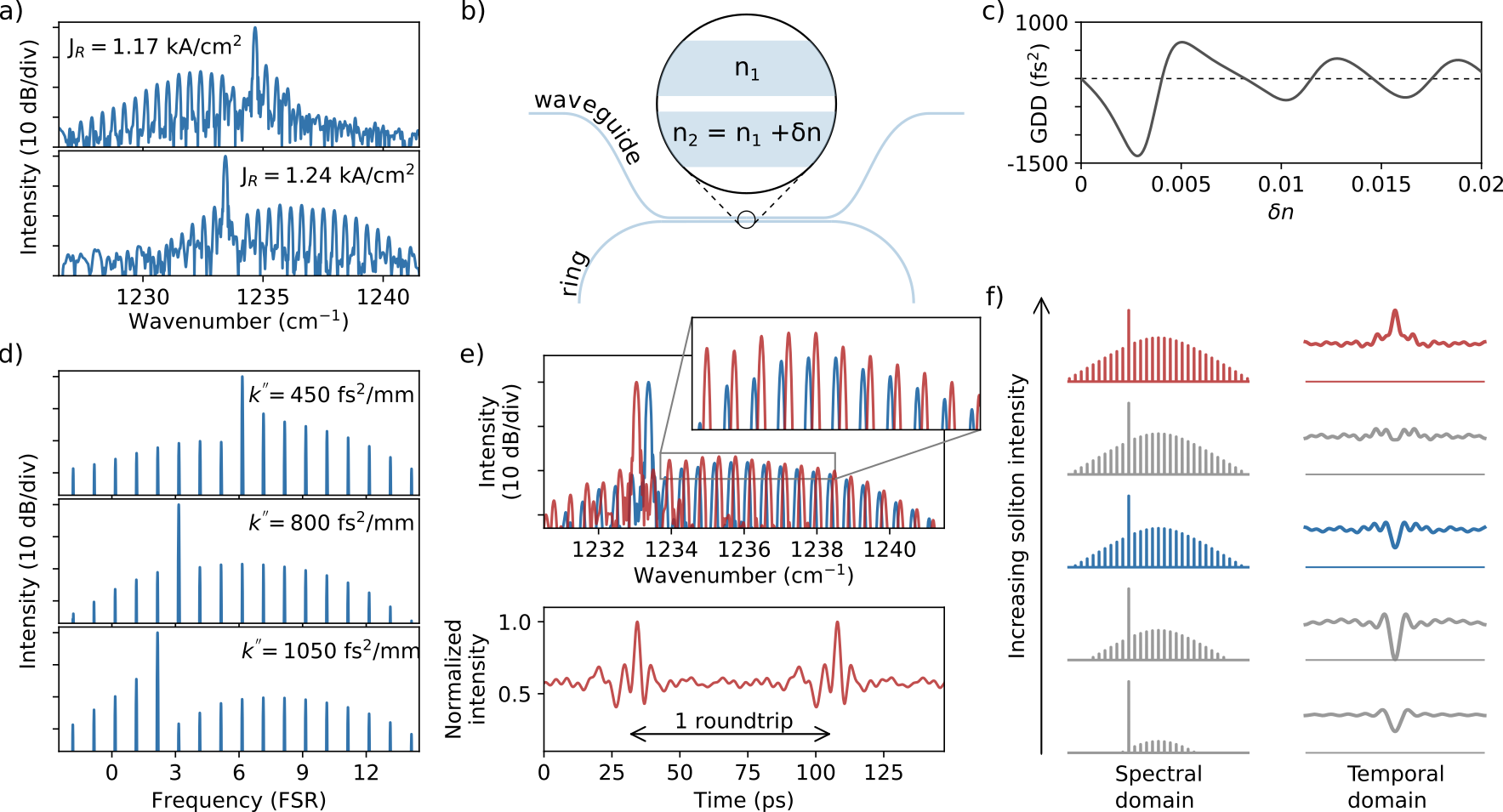}
	\caption{ \textbf{Coherent control of \ac{nb} soliton regimes.} \textbf{a)} Spectral shift of the soliton relative to the primary mode as the ring current is swept. Similar behavior is exhibited if the waveguide current is tuned instead. \textbf{b)} Sketch of the coupled ring-waveguide system. The bias alters the optical indices of the ring and the waveguide through carrier- and thermal-induced changes, resulting in a mismatch (detailed analysis in the Supplementary material). \textbf{c)} Group delay dispersion, calculated with a coupled mode theory analysis, as a function of the index mismatch between the ring and the waveguide. Compared to a single ring cavity, our devices provide superior control of the dispersion. \textbf{d)} Simulated soliton spectra as the cavity dispersion $k^{''}$ is swept, demonstrating a spectral shift of the soliton similar to a). The gainshape contribution to the total \ac{gvd} is around -1500$\,\mathrm{fs}^2/\mathrm{mm}$. \textbf{e)} Experimentally-obtained intensity spectrum of a \ac{nb} soliton at bias currents of $J_{\mathrm{R}}=$1.39$\,\mathrm{kA/cm}^2$ and $J_{\mathrm{WG}}=$0.85$\,\mathrm{kA/cm}^2$ (red), which corresponds to a bright pulse in the intensity waveform. The soliton sidemodes are stronger compared with the dark-pulse soliton from Fig.~\ref{fig2}d) (replotted in blue). \textbf{f)} Illustration of the theoretical dependence of the intensity waveform on the soliton spectrum. We assume ideal \ac{nb} soliotn intermode phase distribution with $\pi$ jumps around the primary mode. }
	\label{fig4}
\end{figure*}

To corroborate the experimental findings, we employ numerical simulations of the master equation derived from the Maxwell-Bloch system (Eq.~(S2) in the Supplementary material)~\cite{opacak2019theory,opacak2021frequency}. 
The obtained comb spectrum, shown in Fig.~\ref{fig2}f), has a smooth spectral envelope engulfing a strong primary mode and spanning over more than 100 cavity modes. 
Numerical simulations are not constrained by a small dynamical range as the experiment (around 35$\, \mathrm{dB}$ in Fig.~\ref{fig2}d)). Hence, by concentrating on the top part of the simulated spectrum within the same range (Fig.~\ref{fig2}g)), we show a clear agreement with the experiment. The simulated intermode phases confirm the $\pi$ shift between the primary mode and the sidemodes, indicating that this is a hallmark of \ac{nb} solitons. The simulated temporal intensity in Fig.~\ref{fig2}h) matches the structures predicted by the \ac{cgle} theory in Fig.~\ref{fig1}. 
Higher-order dispersion due to the laser gainshape is the likely cause of the residual small ringing that trails the \ac{nb} soliton, as is well-known from microresonator solitons~\cite{coen2012modelling,jang2014observation,kippenberg2018dissipative,anderson2022zero}.  
The combined simulated and experimental temporal waveforms confirm that \ac{nb} solitons are surrounded by a \ac{cw} background -- providing a compelling proof of their dissipative soliton nature.

Localization is another striking soliton feature, clearly noticed in multisoliton states -- spontaneously ordered ensembles of several co-propagating solitons~\cite{herr2013temporal,guo2016universal}.  
Multisoliton states can be indubitably identified by their ‘fingerprint’ optical spectra, which have a modulated envelope due to the interference between individual solitons.
Fig.~\ref{fig3}a) depicts one such spectrum, consisted of smooth lobes with spectral holes in-between.
The intermode phases indicate that, besides the usual $\pi$ jumps around the primary mode, an additional $\pi$ jump occurs at the position of 
the spectral hole -- providing another telltale sign of a multisoliton state. 
The reconstructed intensity (Fig.~\ref{fig3}b)) displays two distinct dark pulses,  
during which the phase changes by $4\pi$ -- twice as much as for a fundamental \ac{nb} soliton.
Master equation simulations verify multisoliton states as well. 
The spectral behavior, including the additional $\pi$ intermode phase jump between the lobes, is seen in Fig.~\ref{fig3}c). 
Other than enabling large amplitude contrast, the dynamic range of simulations allows to resolve the $4\pi$ change of the temporal phase in two steps -- one for each soliton (Fig.~\ref{fig3}e)). 
Multistability 
is yet another crucial dissipative soliton trait predicted in Fig.~\ref{fig1}.
To emulate laser starting from spontaneous emission, weak noise from a random number generator with a defined seed is fed as a starting condition into the master equation.  
Changing the seed, while keeping other laser parameters fixed, led to three states with different number of solitons (Fig.~\ref{fig3}f)).
The coexistence of solitons and a single-mode field proves that \ac{nb} solitons are found in the CW linearly-stable region of the parameter space, as predicted by the \ac{cgle}.

Separate electrical contacts of the waveguide and the ring provide two invaluable knobs for soliton control. 
Demonstrating this, we shift the soliton spectral lobe from red to the blue side of the primary mode purely by tuning the ring current in Fig.~\ref{fig4}a).  %, while the waveguide bias is kept constant. 
Besides changing the gain and the \ac{lef}~\cite{opacak2021frequency,piccardo2020frequency}, %through the carrier population 
altering the bias strongly impacts the \ac{gvd}, as confirmed by a coupled mode theory analysis of the ring-waveguide configuration (Figs.~\ref{fig4}b) \& c)).
A small current-induced index mismatch between the ring and the waveguide induces large changes of the \ac{gvd}, covering both normal and anomalous values. 
The key role of \ac{gvd} is also obvious from numerical simulations in Fig.~\ref{fig4}d), where we swept the cavity dispersion. Relative to the primary mode, the soliton is spectrally shifted similarly to Fig.~\ref{fig4}a), and in agreement with recent observations in microresonators~\cite{zhang2022dark}. 
More anomalous \ac{gvd} sets the laser operating point in the linearly-unstable parameter region in analogy to Fig.~\ref{fig1}, where neither single-mode nor \ac{nb} soliton regime can exist. Instead, homoclons form via phase turbulence~\cite{piccardo2020frequency}, which we observe both in experiments and simulations (Supplementary material). 

Having the possibility to form bright pulses with high peak power would open many doors for \ac{nb} solitons to be used in nonlinear processes, such as supercontinuum generation~\cite{obrzud2017temporal}. 
One way of achieving this relies on modifying the phase of the primary mode to remove the destructive interference and induce an intense bright pulse. This could be realized by a second active ring resonator coupled with the waveguide, acting as a notch filter~\cite{liu2021high}. 
Another possibility in this direction is illustrated in
Fig.~\ref{fig4}e), where we show an \ac{nb} soliton spectrum (in red), obtained at a larger bias compared to the soliton from Fig.~\ref{fig2} (replotted in blue). 
Astonishingly, temporal intensity of the former unveils a low-contrast bright pulse.
A comparison between the two soliton spectra reveals that the bright one has stronger sidemodes.
To gain an intuitive understanding, Fig.~\ref{fig4}f) conceptually studies the dependence of the temporal intensity on the corresponding spectrum, while assuming the characteristic intermode phase distribution of \ac{nb} solitons (see Supplementary material for the full analysis). Whereas the primary mode is fixed, the soliton lobe is gradually increased, with the color coding representing the two states from Fig.~\ref{fig4}e). 
The initial increase of the soliton sidemodes enhances the amplitude contrast of a starting dark pulse, until an intensity equilibrium between the primary mode and the sidemodes is reached. 
At this point, destructive interference between the background and the soliton is complete and the dark pulse reaches zero at its minimum. 
Further enhancement of the sidemodes causes a decrease of the pulse amplitude contrast, eventually reaching a quasi-constant waveform despite a broad spectrum, as is also experimentally observed (Supplementary material). Finally, additional sidemode amplification causes the pulse to 'flip over' -- resulting in a bright coherent pulse.
Both the \ac{cgle} and the master equation predict the emergence of \ac{nb} solitons as dark pulses, however neither of the two treatments takes into account the delayed carrier population response to amplitude modulations~\cite{mansuripur2016single}.
A better understanding of the link between the carrier dynamics and the parametric gain, that is necessary for multimode emission, could allow us to optimize active ring resonators for the emission of high-contrast bright pulses.

In this work, we have demonstrated a new way of direct spontaneous soliton generation by utilizing a \ac{mir} semiconductor laser active material implemented in an on-chip integrated ring cavity with a coupler waveguide. 
The waveguide coupler has an independent bias, ensuring not only higher output powers, but also providing a powerful knob to control the total dispersion of the system. Paired with the giant resonant Kerr nonlinearity of the active material, this solidifies the coupled waveguide-ring configuration as a very fruitful playground for nonlinear phenomena -- including the direct generation of electrically-driven \ac{nb} solitons.
The soliton regime is demonstrated by combining both experimental and theoretical results.
The number of solitons within one roundtrip varies stochastically with the initial conditions of the lasers. This is indicative of a multistability phenomenon, typical for dissipative solitons in extended systems, and paves the way for independent soliton addressing~\cite{akhmediev2008dissipative}.
Moreover, we predict that \ac{nb} solitons are not platform-dependent and anticipate their demonstration in other semiconductor laser types, such as the interband cascade- or quantum dot lasers. 
The spontaneous formation of \ac{nb} solitons with current tuning, without the need of an external optical pump, makes our \ac{qcl} rings with coupled waveguides ideal candidates for monolithic soliton generators specifically targeting \ac{mir} applications.

\section*{Methods}
\textbf{Device fabrication and operation}. The lasers emit at around 8.2$\,\mathrm{\upmu m}$ and consist of GaInAs/AlInAs layers on an InP substrate, with the band structure design based on a standard single-phonon continuum depopulation scheme. The waveguides and the narrow gap in the coupling region are etched using the standard fabrication recipe employing optical lithography. The waveguide width is 10$\,\mathrm{mm}$, the curved section of the racetrack is a semicircle with a radius of 500$\,\mathrm{\upmu m}$, and the length of the straight section is 1.5$\,\mathrm{mm}$. The ring circumference defines the cavity repetition rate of around 13.6$\,\mathrm{GHz}$. The heat sink temperature was kept at 14$^\circ$C in all experiments.

\textbf{Experimental setup and characterisation of the soliton states}. The measurement of the spectral amplitudes and phases, that are used for the reconstruction of the temporal waveform, is done by utilizing the SWIFTS technique~\cite{burghoff2015evaluating}. An overview of the measurement procedure is in the Supplementary material.
In order to record the laser output modulation at the repetition frequency, we employ a fast quantum well infrared photodetector in a cryostat, cooled with liquid nitrogen. The setup uses a custom-built high-resolution Fourier transform infrared (FTIR) spectrometer ($\sim$500$\,\mathrm{MHz}$). A Zurich Instruments HF2LI lock-in amplifier is used for the acquisition of both the intensity and SWIFTS interferograms. In order to stabilize the repetition frequency of the comb, we rely on electrical injection locking via a radio-frequency (RF) source. However, caution is required since using a too large injection power will perturb the soliton spectrum and its intermode phases, as can be seen in the Supplementary material.

\textbf{Numerical simulations}.
Simulations of the CGLE are implemented using a pseudo-spectral algorithm coupled with an exponential time-differencing scheme. The master equation of a unidirectional field is discretized with a first-order forward finite difference method. This method hugely benefits from parallel implementation on a graphics processing unit (GPU), cutting down the execution time by 3 orders of magnitude compared to a standard implementation on the central processing unit. As an example, we simulate tens of millions of time steps within a minute using a NVIDIA GeForce RTX 3070 GPU (tens of thousands of simulated cavity roundtrips are necessary to ensure that a steady state is reached). To emulate a laser starting from spontaneous emission, the numerical simulations are run with random weak noise as a starting condition. The latter is obtained from a random number generator by defining an integer seed value.

\section*{Author contributions}
N.~Opa\v{c}ak, D.~Kazakov, F.~Pilat, T.~P.~Letsou, and B.~Schwarz carried out the experiments and analyzed the data. M. Beiser fabricated the device. N.~Opa\v{c}ak performed the master equation simulations. L.~L.~Columbo, M.~Brambilla, and F.~Prati did the \ac{cgle} simulations and contributed to analysis of the experimental results. N.~Opa\v{c}ak prepared the manuscript with input from all coauthors. N.~Opa\v{c}ak and T.~P.~Letsou wrote sections of the Supplementary material. B.~Schwarz, M. Piccardo, and F.~Capasso supervised the project. All authors contributed to the discussion of the results.

\section*{Acknowledgements}
This project has received funding from the European Research Council (ERC) under the European Union’s Horizon 2020 research and innovation programme (Grant agreement No. 853014), and from the National Science Foundation under Grant No. ECCS-2221715. T. P. L. would like to thank the support of the Department of Defense (DoD) through the National Defense Science and Engineering Graduate (NDSEG) Fellowship Program.

\bibliography{darkSol_references.bib}
\clearpage

\begin{acronym}
	
	\acro{mir}[MIR]{mid-infrared}
	\acro{nb}[NB]{Nozaki-Bekki}
	\acro{cw}[CW]{continuous-wave}
	\acro{qcl}[QCL]{quantum cascade laser}
	\acro{fp}[FP]{Fabry-P\'{e}rot}
	\acro{shb}[SHB]{spatial hole burning}
	\acro{cgle}[CGLE]{complex Ginzburg-Landau equation}
	\acro{gvd}[GVD]{group velocity dispersion}
	\acro{lef}[LEF]{linewidth enhancement factor}
	
\end{acronym}

%%%%%%%%%%%%%%%%%%%%%%%%%%%%%%%%%%%%%%%%%%%%%%%%%%%%%%%%%%%%%%%%%%%%%%%%%%%%%%%%%%%%%%%%%%%%%%%%%%%%
%%%%%%%%%%%%%%%%%%%%%%%%%%%%%%%%%%%%%%%%%%%%%%%%%%%%%%%%%%%%%%%%%%%%%%%%%%%%%%%%%%%%%%%%%%%%%%%%%%%%
%%%%%%%%%%%%%%%%%%%%%%%%%%%%%%%%%%%%%%%%%%%%%%%%%%%%%%%%%%%%%%%%%%%%%%%%%%%%%%%%%%%%%%%%%%%%%%%%%%%%
%%%%%%%%%%%%%%%%%%%%%%%%%%%%%%%%%%%%%%%%%%%%%%%%%%%%%%%%%%%%%%%%%%%%%%%%%%%%%%%%%%%%%%%%%%%%%%%%%%%%
%%%%%%%%%%%%%%%%%%%%%%%%%%%%%%%%%%%%%%%%%%%%%%%%%%%%%%%%%%%%%%%%%%%%%%%%%%%%%%%%%%%%%%%%%%%%%%%%%%%%
%%%%%%%%%%%%%%%%%%%%%%%%%%%%%%%%%%%%%%%%%%%%%%%%%%%%%%%%%%%%%%%%%%%%%%%%%%%%%%%%%%%%%%%%%%%%%%%%%%%%
%%%%%%%%%%%%%%%%%%%%%%%%%%%%%%%%%%%%%%%%%%%%%%%%%%%%%%%%%%%%%%%%%%%%%%%%%%%%%%%%%%%%%%%%%%%%%%%%%%%%
%%%%%%%%%%%%%%%%%%%%%%%%%%%%%%%%%%%%%%%%%%%%%%%%%%%%%%%%%%%%%%%%%%%%%%%%%%%%%%%%%%%%%%%%%%%%%%%%%%%%

\renewcommand\appendixname{Supplementary material}
\onecolumngrid
\appendix
\newpage

\renewcommand{\thefigure}{S\arabic{figure}}
\renewcommand{\thetable}{S\arabic{table}}
\renewcommand{\theequation}{S\arabic{equation}}

\section{Master equation for lasers with a fast gain medium}
\label{sec:ME}

Modeling of semiconductor laser multimode dynamics is typically done by employing the Maxwell-Bloch equations. This method, obtained by coupling the density  matrix  formalism with Maxwell’s equations, is fully capable of describing the coherence and the spatio-temporal evolution of the laser light inside the cavity. Within the slowly varying envelope approximation and the rotating wave approximation, which are two commonly-made approximations, the Maxwell-Bloch equations for an open two-level laser system read
\begin{align}
\begin{split}
&\frac{\partial n_{l}}{\partial t}={J_t}+\frac{n_{u}}{T_{lu}}-\frac{n_{l}}{T_{gl}}
-\frac{\mu}{\hbar}\operatorname{Im}\{E\sigma^{\ast}\}, \\
&\frac{\partial n_{u}}{\partial t}=J-\Big(\frac{1}{T_{lu}}+\frac{1}{T_{gu}}\Big)n_{u}
+\frac{\mu}{\hbar}\operatorname{Im}\{E\sigma^{\ast}\},\\
&\frac{\partial \sigma}{\partial t}=-\frac{1+i\alpha}{T_{2}}\sigma
+i\frac{\mu}{2\hbar}(1+i\alpha)^2E(n_{u}-n_{l}), \\
&\Big(\frac{n_r}{c}\frac{\partial}{\partial t}+\frac{\partial}{\partial z}\Big)E=-i\frac{\Gamma \mu \omega_0}{n_r\varepsilon_0 c L_p}\sigma   -\frac{\alpha_w}{2}E.
\label{eq:MBE} 
\end{split}
\end{align}
In the system above, $n_u$ and $n_l$ are the surface carrier densities of the upper and lower laser levels, and $\sigma$ and $E$ stand for the complex slowly-varying envelopes of the off-diagonal density matrix element and the electric field, respectively. Temporal and spatial coordinates are given with $t$ and $z$, and lifetime $T_{ji}$ represents nonradiative transitions from level $i$ to level $j$ (between the ground, the lower, and the upper level). The coherence lifetime is labeled as $T_2$, $\Gamma$ is the light confinement factor of the laser cavity, $\mu$ is the dipole matrix element, $\omega_0$ is the optical frequency of the resonant transition, $n_r$ is the real refractive index, $L_p$ is the thickness of the laser active region, $\alpha_w$ gives the waveguide power losses, the current density $J$ represents the pumping rate to the upper laser level and $J_t$ models the thermal excitation of carriers to the lower lasing level, where both $J$ and $J_t$ are normalized to the elementary charge. Parameter $\alpha$ is the \ac{lef} at the gain peak and it describes the coupling between the amplitude and the phase of the light, which is crucial for semiconductor laser dynamics~\cite{henry1982theory,opacak2019theory}.

The set of equations~(\ref{eq:MBE}) quantitatively describes the spatio-temporal evolution of the laser field. However, following the formalism from~\cite{opacak2019theory}, it is possible to replace the entire system of Maxwell-Bloch equations with a single master equation, thus greatly simplifying the theoretical model an allowing to draw analogies with the \ac{cgle}. The master equation is valid for lasers with a fast gain medium, such as the \ac{qcl}, where the gain recovery time is much shorter than the cavity roundtrip time. The master equation, written for a unidirectional field inside a ring cavity, reads
\begin{align}
\frac{n_r}{c}\frac{\partial E}{\partial t}  + \frac{\partial E}{\partial z} &=\frac{g(I)}{2}  (1+i{\alpha})   \Big(E - \tilde{T}_2\frac{\partial E}{\partial t}    +\tilde{T}_2^2 \frac{\partial^2 E}{\partial t^2}   \Big) 
-\frac{\alpha_w}{2}E   +i\beta{|{E}|}^2E   +  i \frac{k^{''}}{2}\frac{\partial^2E }{\partial t^2} ,
\label{eq:me_unid} 
\end{align}
where $\tilde{T}_2=T_2/(1+i\alpha)$, $k^{''}$ is the cavity \ac{gvd} that originates from the material and waveguide geometry, and $\beta$ is the bulk Kerr nonlinearity coefficient, which we set to zero in this work (a much larger effective Kerr nonlinearity originates from the gain medium of the laser itself). Further, $I=|E|^2$ is the intensity, and $g(I)$ is the saturated gain defined by 
\begin{align}
g(I) = \frac{g_0}{1+\frac{I}{I_s}} ,
\label{eq:sat_gain} 
\end{align}
where the unsaturated gain is $g_0 = \Gamma\mu^2\omega_0T_1T_2J/(\hbar n_r c \varepsilon_0 L_p)$, the saturation intensity is $I_s=2\hbar^2/(\mu^2T_1 T_2)$, and the carrier recombination time is $T_1=(T_{lu}^{-1}+T_{gu}^{-1})^{-1}$ (approximately equal to the gain recovery time). 
The complete derivation, starting from the Maxwell-Bloch system, can be found in detail in~\cite{opacak2022thesis}. 

\subsection{The total GVD and Kerr nonlinearity in the laser}
\label{sec:disucssion_parameter}

It is noteworthy to underline that the total \ac{gvd} in the laser consists of two major contributions. The first, proportional to the coefficient $k^{''}$ in the model, describes the dispersion that comes from the bulk material of choice and also the mode dispersion due to the resonator geometry. We refer to it as the cavity dispersion and, as an approximation, it is considered to be a constant pure second-order dispersion.
The second major contribution to the dispersion originates from the the resonant optical transition itself, and we refer to it as the gain dispersion. It is inherently included in the Maxwell-Bloch system of equations~(\ref{eq:MBE}) in the equation for the off-diagonal matrix $\sigma$. 
This equation describes the response of the material to the electric field, captured by the complex susceptibility in the spectral domain
\begin{align}
\chi(\omega) = \frac{\mu^2(n_u-n_l)}{\varepsilon \hbar L_p}\frac{(1+i\alpha)^2}{\omega-\omega_0-\frac{i}{T_2}},
\label{eq:chi} 
\end{align}
where $\operatorname{Im}\{\chi\}$ models the modified Lorentzian spectral lineshape of the gain (ideal Lorentzian shape for vanishing \ac{lef} $\alpha=0$).
Moreover, the complex susceptibility also describes the spectral dependence of the refractive index $n_r(\omega)\approx\sqrt{1+\chi_{\mathrm{bulk}}+\operatorname{Re}\{ \chi(\omega)\}}$. Keeping in mind the definition $\mathrm{GVD}(\omega)=\frac{2}{c}\frac{\partial n_r}{\partial \omega}+\frac{\omega}{c}\frac{\partial^2 n_r}{\partial \omega^2}$, it becomes obvious how the resonant optical transition induces an additional gain dispersion. This contribution to the total \ac{gvd} includes also higher-order terms besides the second-order dispersion, however, its approximate value at position of the gain peak is given in relation~(\ref{eq:gainGVD}). 

The \ac{lef} physically models the coupling between the amplitude and phase of the light, and it is mathematically defined as the ratio of the carrier-induced changes of the refractive index and the optical gain~\cite{henry1982theory}.
As such, the \ac{lef} describes how the asymmetric spectral shape of $\operatorname{Im}\{\chi\}$ alters the value of $\operatorname{Re}\{\chi\}$ at the gain peak and shifts it away from zero ($\operatorname{Re}\{\chi\}$ has a zero-crossing at the gain peak for an ideal symmetric Lorentzian shape of $\operatorname{Im}\{\chi\}$). This physically means that whenever the carrier population changes, a corresponding variation of not only the gain, but also of the refractive index is induced. 
In a laser system, the carriers are modulated by the propagating light inside the cavity, which results in population pulsations due to dynamic gain saturation~\cite{mansuripur2016single}. 
As a result, any change of the light intensity indirectly modifies the refractive index of the medium through these population pulsations. 
This is tightly related to the appearance of a Kerr nonlinearity~\cite{opacak2019theory,opacak2021frequency}. Its origin is resonant, as it comes from the gain medium itself and is relevant for optical frequencies within the gainwidth of the laser. This contrasts with the broadband bulk nonlinearity, which is proportional to the coefficient $\beta$ in the model.   
The above-presented arguments are valid assuming that the population pulsations can follow the speed of the intensity modulations, which occur at the cavity repetition frequency. In other words, in media with fast gain recovery times, a finite value of the \ac{lef} phenomenologically induces a giant resonant Kerr nonlinearity, whose value is approximately given by equation~(\ref{eq:kerrGain}).

\begin{table}
	\begin{tabular}{ |p{1.5cm}||p{7cm}||p{2.5cm}|  }
		%\hline
		%\multicolumn{3}{|c|}{Parameters} \\
		\hline
		Symbol& Description & Value\\
		\hline
		$T_{lu}$ & Upper-lower level transition lifetime   & $1.15 \: \mathrm{ps} $ \\
		$T_{gu}$ & Upper-ground level transition lifetime     & $6 \: \mathrm{ps} $ \\
		$T_{gl}$ & Lower-ground level transition lifetime     & $80 \: \mathrm{fs} $ \\
		$T_{2}$ & Dephasing time  & $125 \: \mathrm{fs} $ \\
		$n_r$ & Refractive index  & $3.37 $ \\
		$\alpha_{w}$ & Waveguide power losses   & $4 \: \mathrm{cm}^{-1} $ \\
		$\mu$ & Dipole matrix element  & $1.8 \: \mathrm{nm}\times\mathrm{e}  $ \\
		\hline
	\end{tabular}
	\caption{\label{demo-table} Parameter values used in the master equation.}
\end{table}

\section{Complex Ginzburg-Landau equation for ring semiconductor lasers}
\label{sec:CGLE}

The dynamics of ring semiconductor lasers with a fast gain medium close to the lasing threshold are well captured by the the \ac{cgle}~\cite{aranson2002world,piccardo2020frequency}. Here we will start from the more general master equation~(\ref{eq:me_unid}) and derive the \ac{cgle}, similarly to the procedure in Refs.~\cite{piccardo2020frequency,columbo2021unifying}.
In the first step, the term proportional to $\tilde{T}_2 \partial E/\partial t$ on the right hand side of the master equation~(\ref{eq:me_unid}) is neglected.
This is justified, as its main effect is to introduce a constant shift of the value of the refractive index $n_r$. 
Additionally, substituting the second time derivative $\partial^2/ \partial t^2$ with the second spatial derivative $ (c/n_r)^2 \partial^2/ \partial z^2$ is an excellent approximation, as the equation deals with slowly varying envelopes in the rotating wave approximation.  Moreover, the system is switched to a frame of reference that moves together with the propagating field by applying coordinate transformations $z \rightarrow z - \frac{c}{n_r}t$ and $t \rightarrow t$. Equation~(\ref{eq:me_unid}) then transforms to
\begin{align}
\frac{n_r}{c}\frac{\partial E}{\partial t} & =\frac{g(I)}{2}  (1+i{\alpha})   \Big(E    +\tilde{T}_2^2 \frac{c^2}{n_r^2} \frac{\partial^2 E}{\partial z^2}   \Big) 
-\frac{\alpha_w}{2}E   +i\beta{|{E}|}^2E   +  i \frac{k^{''}}{2}\frac{c^2}{n_r^2}\frac{\partial^2 E }{\partial z^2} .
\label{eq:gl1} 
\end{align}
We can use the Taylor expansion to approximate the saturated gain $g(I)$ in the vicinity of the stationary intensity $I_0=I_s(\frac{g_0}{\alpha_w}-1)$
\begin{align}
g(I) = \frac{g_0}{1+ \frac{I}{I_s}} \approx g(I_0) + \frac{\partial g(I)}{\partial I} \big|_{I=I_0} (I-I_0) \approx g_1 - g_2 \frac{I}{I_s} ,
\label{eq:gl2} 
\end{align}
where $g_1 = \frac{\alpha_w}{g_0}(2g_0-\alpha_w)$ and  $g_2 = \frac{\alpha_w^2}{g_0}$. Inserting relation~(\ref{eq:gl2}) into equation~(\ref{eq:gl1}) and neglecting the term proportional to $\sim |E|^2\partial^2E/\partial z^2$ results in
\begin{align}
\begin{split}
\frac{\partial E}{\partial t}  = (\eta + i \omega_s)E + (d_R+id_I) \frac{\partial^2E}{\partial z^2} - (n_R+in_I)|E|^2E ,
\label{eq:gl8} 
\end{split}
\end{align}
where the following functions have been introduced
\begin{align}
\begin{split}
&\eta = \frac{g_1-\alpha_w}{2} \frac{c}{n_r},  \: \: \:\:\:\:\: \: \:\:\: \omega_s = \frac{g_1{\alpha}}{2}\frac{c}{n_r}, \\
& d_R= \frac{g_1 T_2^2}{2(1+{\alpha}^2)} \frac{c^3}{n_r^3}, \: \: \:\:\:\: d_I = \Big( \frac{k^{''}}{2} - \frac{{\alpha}g_1T_2^2}{2(1+{\alpha}^2)} \Big) \frac{c^3}{n_r^3}  \\
& n_R = \frac{g_2}{2 I_s} \frac{c}{n_r},  \: \: \:\:\:\: \: \: \:\:\:\:\: \: \:\:\:  n_I= -\Big(  \beta  -\frac{{\alpha} g_2}{2 I_s} \Big)\frac{c}{n_r}.
\label{eq:gl9} 
\end{split}
\end{align}
Here, the laser net gain is described with the coefficient $\eta$, and $\omega_s$ represents the frequency shift due to the gain asymmetry quantified by ${\alpha}$ (value of \ac{lef} at the gain peak). The complex diffusion coefficient, given with $d_R+id_I$, dampens variations of the field $E$. Its complex value can be easily understood as it will work towards smoothing any
spatial gradient of both the amplitude and phase. Physically, it emerges from the curvature of the gain (due to its finite bandwidth), asymmetric gainshape, and the \ac{gvd}. Lastly, the nonlinearity is described with $n_R+in_I$. The real part $n_R$ arises from the gain saturation in the laser and dampens amplitude fluctuations. The imaginary part $n_I$ describes the phase modulation due to the bulk Kerr nonlinearity and a finite \ac{lef}.

Equation~(\ref{eq:gl9}) can be written in a more elegant way. To do so, the temporal, spatial variable, and the electric field need to be rescaled $t \rightarrow \frac{t}{\eta}$, $ z \rightarrow \Big(\frac{d_R}{\eta}\Big)^{1/2} z $, and $ E \rightarrow  \Big(\frac{\eta}{n_R}\Big)^{1/2} e^{i \omega_s t/\eta}E$ .
Moreover, the dispersive and nonlinear parameters are introduced as $c_{\mathrm{D}}$ and $c_{\mathrm{NL}}$ in the following way
\begin{align}
\begin{split}
& c_{\mathrm{D}} =-{\alpha} + k^{''}\frac{g_0(1+{\alpha}^2)}{\alpha_w(2g_0-\alpha_w) T_2^2}, \\
& c_{\mathrm{NL}} = {\alpha} - \beta\frac{2 I_sg_0}{\alpha_w^2}.
\label{eq:gl11} 
\end{split}
\end{align}
From the previous relation, it is clear how the cavity dispersion $k^{''}$ influences $c_{\mathrm{D}}$ and the bulk Kerr nonlinearity coefficient $\beta$ influences $c_{\mathrm{NL}}$, while the \ac{lef} influences both of them, providing a physical contribution to both the ive and nonlinear effects. Equation~(\ref{eq:gl11}) provides also a straightforward way to quantify gain dispersion (see subsection~\ref{sec:disucssion_parameter}) as 
\begin{align}
\mathrm{GVD}_{\mathrm{gain}}= -\frac{\alpha}{1+\alpha^2} \frac{(2g_0-\alpha_w)\alpha_w T_2^2}{g_0},
\label{eq:gainGVD}
\end{align}
and also a giant resonant Kerr nonlinearity that arises from the gain, phenomenologically described with \ac{lef} as
\begin{align}
{\beta}_{\mathrm{gain}}= \alpha \frac{\alpha_w^2 }{2g_0 I_s}.
\label{eq:kerrGain}
\end{align}
Implementing relation~(\ref{eq:gl11}) into equation~(\ref{eq:gl9}) allows us to obtain the conventional form of the \ac{cgle} often found in literature~\cite{aranson2002world}
\begin{align}
\frac{\partial E}{\partial t}  = E + (1+ic_{\mathrm{D}}) \frac{\partial^2E}{\partial z^2} - (1+ic_{\mathrm{NL}})|E|^2E .
\label{eq:cgle} 
\end{align}
In comparison with the laser master equation~(\ref{eq:me_unid}), the entire parameter space of \ac{cgle} has been reduced to just two dimensions defined by the $c_{D}-c_{NL}$ plane.
The \ac{cgle} represents one of the most known nonlinear equations in physics. It generally describes the dynamics of oscillating, spatially extended systems going through phase transitions -- covering a vast amount of nonlinear phenomena on a qualitative, and often on a quantitative level.

%\subsection{NB soliton solutions of the CGLE}
%\label{sec:nbsol_in_cgle}

\section{SWIFTS measurement and experimental characterization of the NB soliton}
\label{sec:swifts}

Here we explain briefly the working principle of the \ac{swifts}~\cite{burghoff2015evaluating}, which is used to experimentally characterize \ac{nb} solitons. \ac{swifts} uses a linear optical intensity detector with a large electrical bandwidth
to detect the phase and amplitude of the intermode beatings of the frequency comb equidistant modes. The optical frequency and phase of a comb mode $n$ is given by $f_n = nf_{\mathrm{rep}}+f_{\mathrm{ceo}}$
and $\phi_n$, respectively ($f_{\mathrm{ceo}}$ is the carrier envelope offset frequency, and $f_{\mathrm{rep}}$ is the repetition frequency). Each pair of adjacent modes
beats at their difference frequency, which is equal to $f_{\mathrm{rep}}$. In order to stabilize $f_{\mathrm{rep}}$ against fluctuations or drifting, we employ electrical injection locking to an external \ac{rf} source. The phase of the intermode beating is equal to the phase difference of the adjacent modes $\phi_n-\phi_{n-1}$. The temporal waveform of the frequency comb can be reconstructed by extracting its intensity spectrum and the phases of all intermode beatings. In order to measure the phases of the individual intermode beatings, a method to isolate pairs of adjacent comb lines, i.e. a frequency discriminator, is required.
\ac{swifts} implements this function in the following way: the light emitted by the frequency comb passes through a \ac{ftir} spectrometer as the delay arm is swept and is subsequently detected by a fast \ac{qwip}. Here,
the \ac{ftir} spectrometer acts as a frequency discriminator in the spectral
domain. The beatnote signal at the repetition frequency, detected by the \ac{qwip} after the \ac{ftir} spectrometer, is downmixed to about 10$\,\mathrm{MHz}$ with a \ac{lo}. The quadratures of the downmixed beatnote signal, $X$
and $Y$, are recorded by a lock-in amplifier as function of the \ac{ftir} interferometer path difference. The reference signal for the lock-in amplifier is provided by the \ac{rf} source, to which the comb repetition rate is locked. This allows to record
the so-called \ac{swifts} interferograms for both quadratures. The
DC signal of the \ac{qwip} yields the intensity interferogram
from which the intensity spectrum of the laser output is obtained via Fourier transform.
A Fourier transform of the \ac{swifts} interferograms yields the complex \ac{swifts} spectrum, where the amplitude of the mode $S_n$ is equal to the geometric average of the amplitudes of adjacent modes in the intensity spectrum $\sqrt{I_nI_{n-1}}$, provided that the characterized state is a coherent frequency comb. If phase noise is present in the part of the spectrum, or the modes are not equidistant, then that part of the  \ac{swifts} spectrum will be substantially weaker. As a conclusion,  comparing the envelopes of the \ac{swifts} and intensity spectra allows to assess the phase coherence of the multimode field and benchmark frequency comb operation.

\begin{figure*}[h]
	\centering
	\includegraphics[width = 1\textwidth]{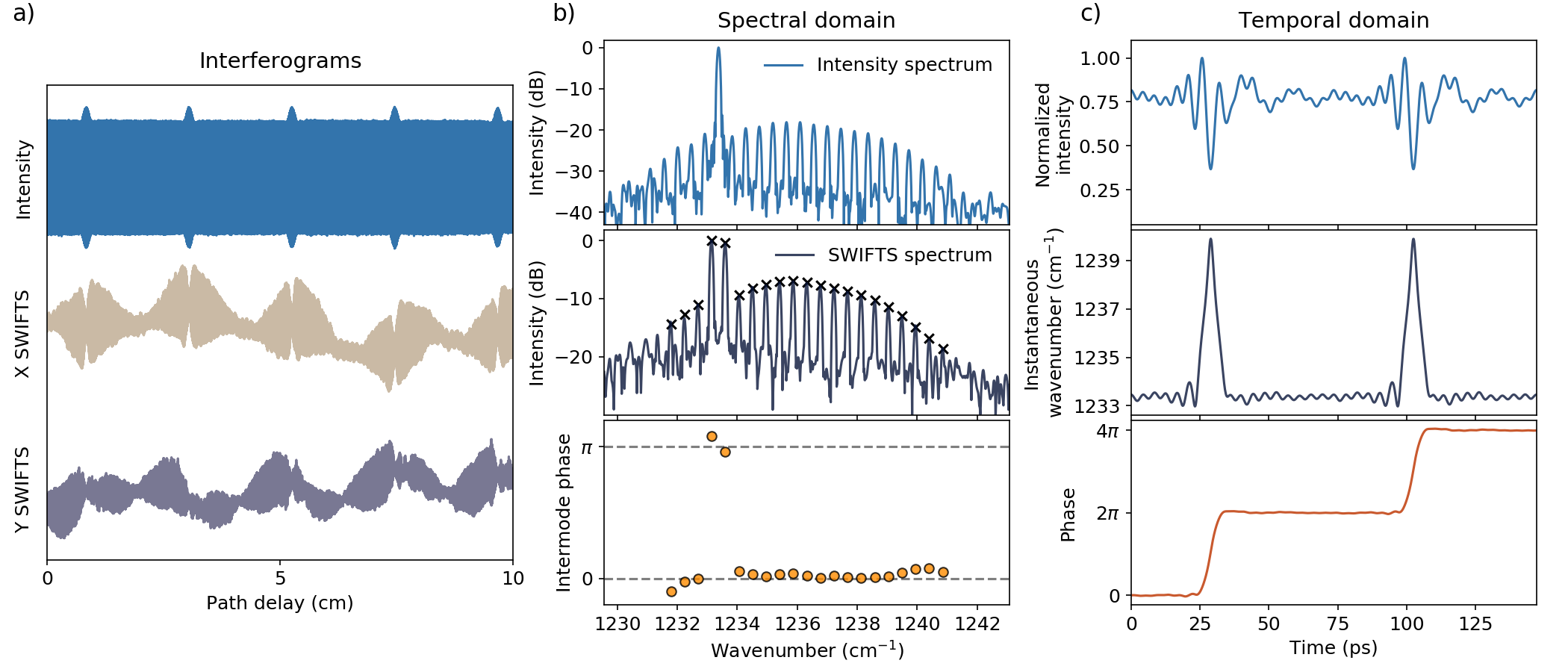}
	\caption{ \textbf{Experimental characterization of a \ac{nb} soliton.} \textbf{a)} Intensity and quadrature $X$ and $Y$ SWIFTS interferograms. \textbf{b)} Intensity spectrum (top), SWIFTS spectrum (middle), and the corresponding intermode phases (bottom). The agreement between the SWIFTS modal amplitudes and the geometric average of neighboring intensity modes (black markers) indicates that the characterized state is coherent and that the laser emits a frequency comb. Intermodal phases show that all of the intermode beatings are synchronized in-phase, except for the two beatings between the primary mode and its neighboring sidemodes, meaning that there is a $\pi$ shift in the phase of the primary mode. \textbf{c)} The reconstructed temporal profiles of the intensity (top), instantaneous wavenumber (middle), and the phase (bottom). The instantaneous wavenumber changes only within the region of the dark pulse, where the phase exhibits a steep ramp covering the value of $2\pi$. The phase is perfectly linear elsewhere, which corresponds to a constant instantaneous wavevector of around $1233.5\,\mathrm{cm}^{-1}$.}
	\label{sup_nb}
\end{figure*}

Fig.~\ref{sup_nb} displays in detail the measured \ac{swifts} data and the characterization of the \ac{nb} soliton from Fig.~2 of the main manuscript (obtained at bias currents of $J_{\mathrm{R}}=$1.33$\,\mathrm{kA/cm}^2$ and $J_{\mathrm{WG}}=$0.79$\,\mathrm{kA/cm}^2$). The intensity interferogram and both of \ac{swifts} quadrature interferograms can be seen in Fig.~\ref{sup_nb}a) as the delay arm of the \ac{ftir} spectrometer is swept. The intensity interferogram has a unique shape characteristic to pulsed laser emission where a strong \ac{cw} background is present. This can be recognized from the interferogram envelope that exhibits pulse-like shapes superimposed on a constant background. The pulses occur approximately every 2.25$\,\mathrm{cm}$ in the path delay, which corresponds to the distance that the light traverses in air during one laser cavity roundtrip. The information about the spectral intermode phases is contained in the $X$ and $Y$ \ac{swifts} interferograms. Both of them have minima at the zero path delay (where the intensity interferogram exhibits pulses), strongly indicating that not all of the intermode beatings are synchronized in-phase. 
We show the spectral behavior of the \ac{nb} soliton in Fig.~\ref{sup_nb}b). The intensity spectrum (top) displays a strong primary mode surrounded by a multitutde of weaker sidemodes that form a smooth bell-shaped envelope, akin to typical dissipative soliton spectra found in passive microresonators or fibers~\cite{herr2013temporal,kippenberg2018dissipative,herr2012universal,guo2016universal,xue2015mode,zhang2022dark,rowley2022self,leo2010temporal,englebert2021temporal}. The modes of the \ac{swifts} spectrum $S_n$ (middle) correspond to the intermode beatings between adjacent optical modes $n$ and $n-1$. The markers indicate the geometric average of the intensity of these neighboring modes, mathematically written as $\sqrt{I_nI_{n-1}}$. From the figure, we can see that the \ac{swifts} modal amplitudes match the values given by the markers, thus proving coherent frequency comb operation. The phases of the complex modes in the \ac{swifts} spectrum correspond to the intermode phases between adjacent modes of the intensity spectrum ($\phi_n-\phi_{n-1}$). The intermode phase distribution, shown in the bottom of Fig.~\ref{sup_nb}b), indicates that all of the intermode beatings are synchronized in-phase, except for the two beatings between the strong primary mode and its neighbors, which are $\pi$-shifted. If translated to the modal phases $\phi_n$, the phases of the sidemodes are approximately identical (in the general case, an arbitrary linear distribution), with the exception of the primary mode, which is $\pi$ out of phase.  Intuitively, this means that the sidemodes interfere constructively in a localized portion of the cavity roundtrip, while the strong primary mode  destructively interferes with them. The knowledge of the intermode phases allows us to extract the phases of the individual modes $\phi_n$ via a cumulative sum. Together with the modal amplitudes, this enables the temporal reconstruction of the electric field as 
\begin{align}
E(t)=\sum_n \sqrt{I_n}\mathrm{cos}(2\pi f_n t + \phi_n).
\label{eq:sum}
\end{align}
We show the temporal intensity of the \ac{nb} soliton in the top of Fig.~\ref{sup_nb}c). It demonstrates a clear dark pulse, which is a consequence of the above-mentioned destructive interference. The pulse is surrounded by a quasi-constant intensity which exhibits residual smaller oscillations. These arise due to a finite number of elements in the Fourier sum in equation~(\ref{eq:sum}), which is a consequence of a limited dynamic range of detection of the \ac{qwip}.
Examining the plotted instantaneous frequency of the output (proportional to the instantaneous wavenumber) is beneficial for understanding the salient properties of these waveforms. It is observed that the frequency is swept only within the region of the dark pulse, indicating that all of the optical modes contribute only within this portion of the roundtrip. The instantaneous frequency is constant everywhere else and equal to the optical frequency of the primary mode (wavenumber $\sim 1233.5\,\mathrm{cm}^{-1}$). This means that the constant portion of the intensity is indeed comprised of a single-frequency \ac{cw} field -- indicating the localized soliton nature of the pulse. 
From the point of view of the one-dimensional \ac{cgle}, the dark pulses can be viewed as sources that connect two plane waves with identical wavenumbers, consistent with periodic ring boundary conditions~\cite{aranson2002world,bekki1985formations,lega2001traveling}. These structures are found to be dynamically stable solutions of the \ac{cgle} in a finite, but narrow region of the $c_{\mathrm{D}}$--$c_{\mathrm{NL}}$ parameter space. In the case when higher-order terms are included, e.g. in a quintic \ac{cgle} with a fifth-order nonlinearity, a co-propagating hole-shock solutions still persist as stable structures in the waveform under ring boundary conditions~\cite{popp1993localized,popp1995hole}. 

The temporal phase profile in Fig.~\ref{sup_nb}c)  exhibits a steep ramp that covers the value of $2\pi$ within the width of the pulse, as discussed in the main manuscript. Together with the $\pi$ jumps of the intermode phases around the primary mode, this represents a 'fingerprint' trait of \ac{nb} solitons that can be used for their unequivocal classification.

\subsection{Spectrally shifted NB solitons}
\label{sec:shifted_sol}

Here we present experimental states with the \ac{nb} soliton spectral envelope shifted relative to the position of the primary mode, as reported in Fig.~4a) from the main manuscript.

\begin{figure*}[t]
	\centering
	\includegraphics[width = 0.6\textwidth]{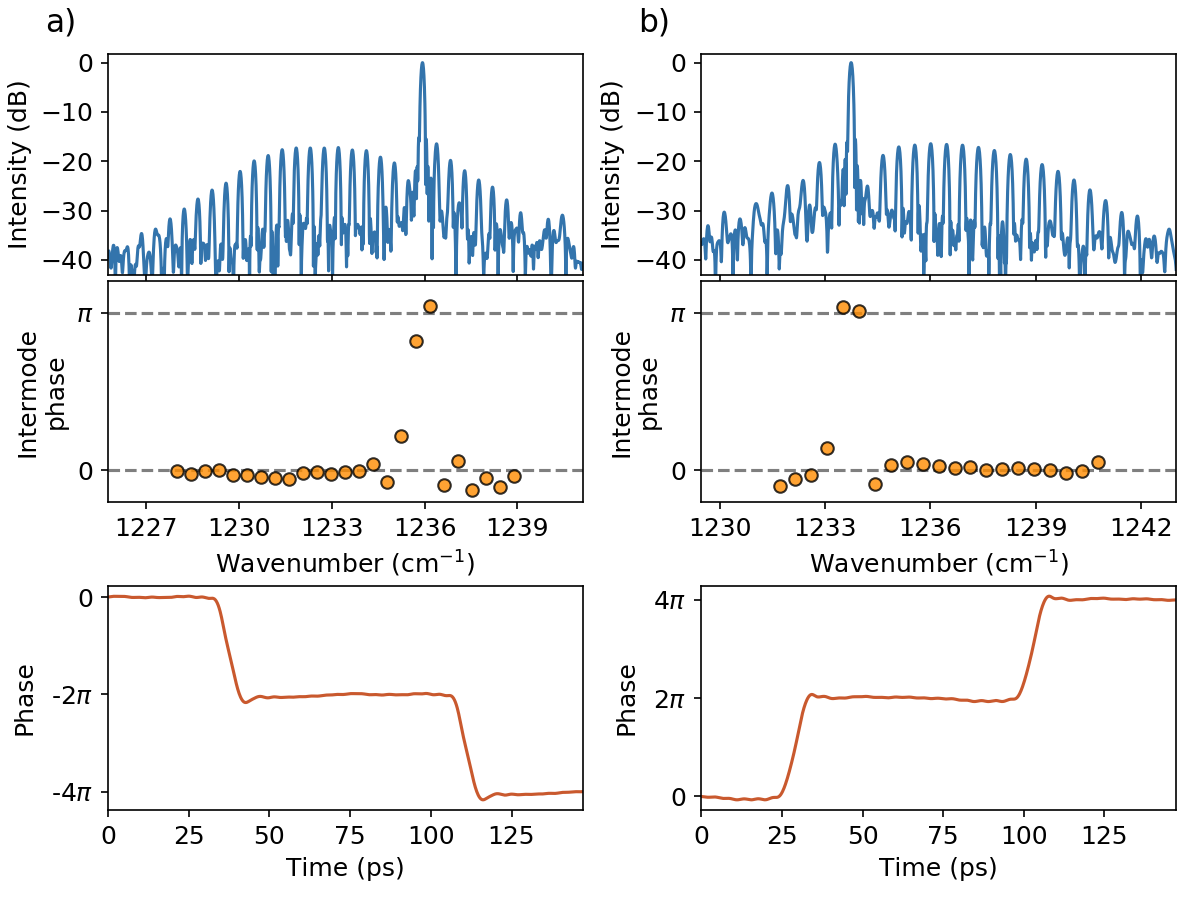}
	\caption{ \textbf{Shifting of the soliton spectral envelope relative to the primary mode.} Experimental characterization of two \ac{nb} solitons where the tuning of the bias current results in a shift of the spectral soliton envelope from the red to the blue side of the primary mode (\textbf{a)} and \textbf{b)} respectively). The intermode phases display the expected two $\pi$ jumps around the primary mode. The temporal phase profile sweeps over $2\pi$ during the soliton width. However, the direction of the sweep depends where the soliton envelope is positioned relative to the primary mode.}
	\label{sup_shift}
\end{figure*}

The shift of the soliton spectral envelope in Fig.~\ref{sup_shift} happens as the currents of the ring and the waveguide are changed. The main reason for this likely lies in the large change of the total \ac{gvd}, as discussed in the main manuscript, and recently observed experimentally in passive microresonators~\cite{zhang2022dark}. 
Although the soliton envelope may be positioned differently relative to the primary mode, the expected two $\pi$ jumps of the intermode phases around the primary mode are still present -- indicating that this is indeed a salient feature of \ac{nb} solitons.

The temporal profile of the phase exhibits the familiar $2\pi$ ramp within the width of the soliton. We can observe that the direction of the ramp depends on whether the soliton spectral envelope is on the red or on the blue side relative to the primary mode. In a hypothetical state where the soliton envelope would be perfectly symmetric relative to the primary mode (if the soliton spectral center of mass coincides with the position of the primary mode), the $2\pi$ phase ramp would comprise two separate $\pi$ ramps with an opposite direction. However, this state does not represent a stable fixed point, and is likely never to occur experimentally.

\section{Experimental evidence of an NB soliton crystal}
\label{sec:crystal}

In the Fig.~3 of the main manuscript, we have shown an experimental and theoretical characterization of a multisoliton state comprised of two co-propagating \ac{nb} solitons in a single roundtrip.
A special case of multisolitonic states, where all of the solitons within one roundtrip are equidistant, is called a soliton crystal~\cite{karpov2019dynamics}. In the frequency domain, these waveforms correspond to a harmonic frequency comb whose spacing between adjacent comb modes is equal to an integer multiple of the \ac{fsr}: $\mathrm{N}\: \times \: \mathrm{FSR}$, where $\mathrm{N}$ is the number of solitons in the soliton crystal.

\begin{figure*}[t]
	\centering
	\includegraphics[width = 0.45\textwidth]{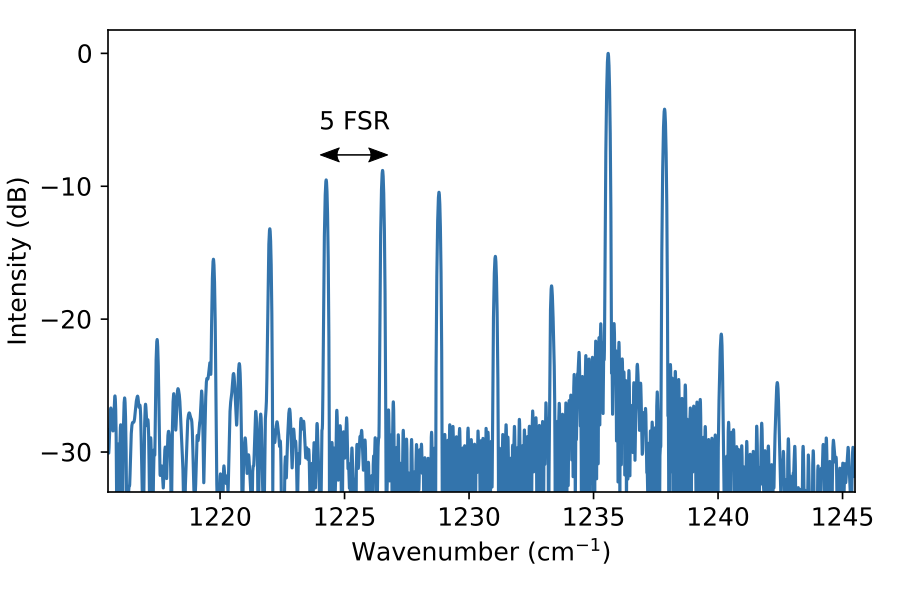}
	\caption{ \textbf{Experimental evidence suggesting an NB soliton crystal.} Intensity spectrum of a probable fifth harmonic frequency comb, where the intermode spacing equals 5 FSRs. The soliton crystal regime is suggested by the smooth bell-shaped envelope of the spectrum. }
	\label{sup_crystal}
\end{figure*}

A ring resonator employing a laser active region can provide a platform for studying soliton crystal dynamics as well.
In Fig.~\ref{sup_crystal}, we report a state with an intensity spectrum whose intermode spacing is equal to 5 \acp{fsr}, potentially corresponding to five equidistant co-propagating \ac{nb} solitons inside the ring cavity. The coherence of the state is suggested by the high suppression ratio of the fundamental modes that fall beneath the noise floor, leaving only the harmonic equidistant modes. Furthermore, the modes form a smooth bell-shaped spectral envelope that indicates the soliton nature of the state.
The high frequency of the intermode beatnote (around 68$\,\mathrm{GHz}$) lies well above the cutoff frequency of our optical detector, thus prohibiting \ac{swifts} characterization to truly assess the coherence of the state. This begs for the future use of another coherent technique to study soliton crystal dynamics in active ring resonators.

\section{Impact of strong radio-frequency injection}
\label{sec:rf}

In order to stabilize the fundamental frequency comb repetition frequency (around 13.6$\,\mathrm{GHz}$) and perform the \ac{swifts} measurement, we employ electrical injection locking via an \ac{rf} source. To do so, injected power below 0$\,\mathrm{dBm}$ is sufficient for our devices.
Going beyond that level and injecting stronger power, although beneficial for locking the comb repetition frequency to the external source, perturbs the salient characteristics of free-running \ac{nb} solitons.
For that reason, we avoid using large \ac{rf} injection powers.

\begin{figure*}[h]
	\centering
	\includegraphics[width = 0.7\textwidth]{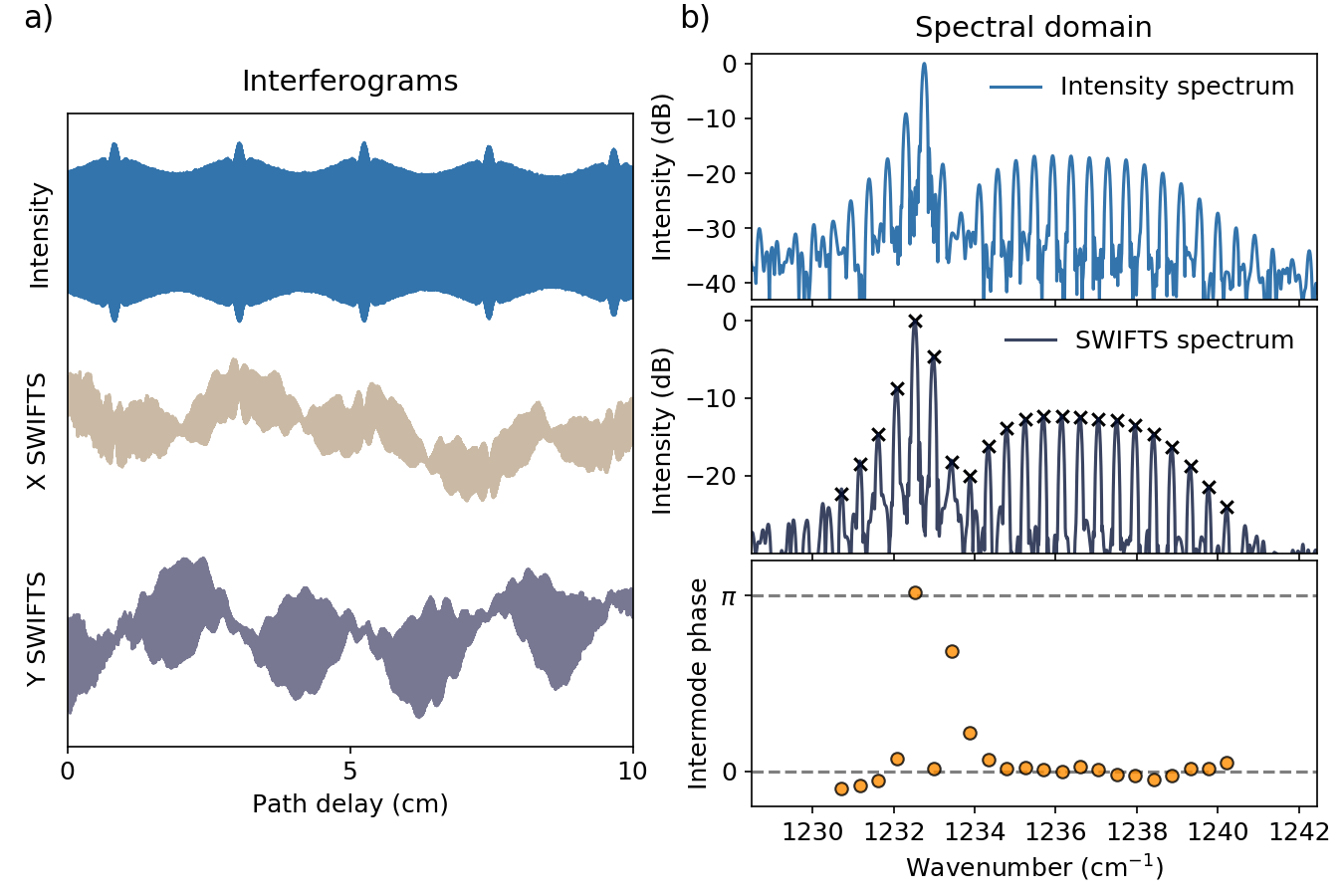}
	\caption{ \textbf{Influence of strong RF injection on the \ac{nb} soliton.} Measured \textbf{a)} intensity and \ac{swifts} interferograms, and \textbf{b)} the intensity spectrum, \ac{swifts} spectrum, and the intermode phases. The intensity interferogram exhibits a modulation of the envelope due to the strong \ac{rf} injection. The intermode phases are affected and the two $\pi$ jumps around the primary mode cannot be seen.   } 
	\label{sup_RF}
\end{figure*}

The influence of strong \ac{rf} injection on an \ac{nb} soliton can be seen in Fig.~\ref{sup_RF}. The first indication that the electrical injection perturbs the free-running soliton can be observed from the intensity interferogram. Unlike the flat interferogram envelope in Fig.~\ref{sup_nb}, we can see that strong \ac{rf} injection causes approximately sinusoidal modulation of the interferogram envelope due to the fact that the strong primary mode develops modulation sidebands. Furthermore, the intermode phase distribution is affected around the primary mode, and the 'fingerprint' two $\pi$ jumps are not present anymore.

\section{Homoclons in the theory and experiment}
\label{sec:homoclons}

In this section, we present theoretical and experimental results of homoclons in our lasers, with the aim of distinguishing these states from \ac{nb} solitons.
Homoclons represent stable frequency combs, characterized by the formation of localized patterns in the waveform~\cite{aranson2002world}. 
They arise due to phase turbulence and have been observed in Ref.~\cite{piccardo2020frequency}, which triggered the subsequent research of ring \ac{qcl} comb dynamics.
They have been observed in ring \acp{qcl}, where they arise due to phase turbulence~\cite{piccardo2020frequency}.%, as shown in Ref.~\cite{piccardo2020frequency}. 

\begin{figure*}[h]
	\centering
	\includegraphics[width = 1\textwidth]{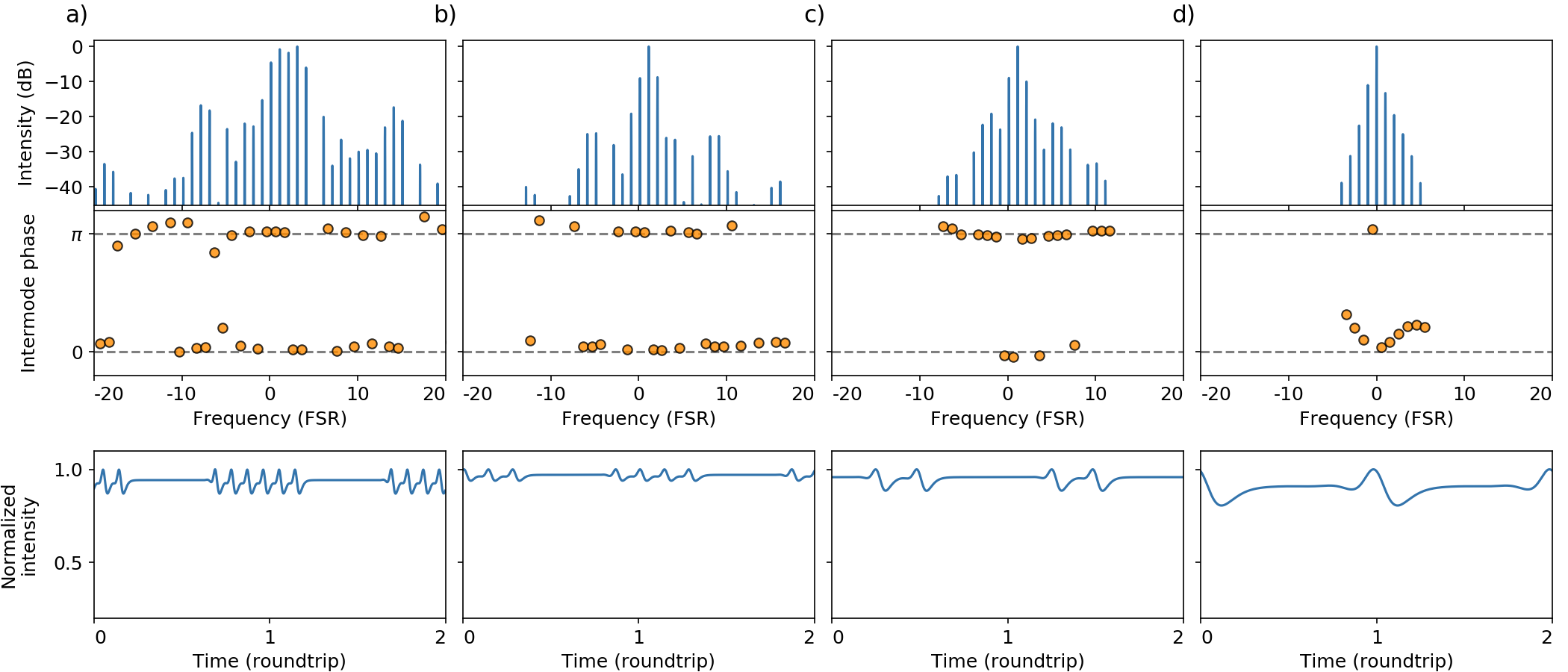}
	\caption{ \textbf{Simulated homoclon states.} \textbf{a)} - \textbf{d)} Exemplary homoclons obtained with numerical simulations of the master equation~(\ref{eq:me_unid}). Homoclons are characterized as localized states that comprise several pulse-like features in the temporal intensity. The number of these features depends on the parameters, but also on the starting conditions of the laser. The intermode phases are distributed in two clusters, separated by $\pi$. The case of one pulse-like feature in column d), cannot be mistaken with an \ac{nb} soliton, as the latter always has two $\pi$ jumps around the primary mode. } 
	\label{sup_theoryHomo}
\end{figure*}

Fig.~\ref{sup_theoryHomo} displays four different homoclon states obtained from numerical simulations of the master equation. The temporal intensity profile exhibits localized patterns that comprise different number of pulse-like structures. Their number depends on the laser parameters, but also on the laser starting condition (phenomenon of multistability). Typical values of \ac{lef} that are used to obtain homoclons are in the range of 1.1-1.3. The cavity dispersion $k^{''}$ falls between -1000 and -1500$\,\mathrm{fs}^2/\mathrm{mm}$, which puts the value of the total \ac{gvd} between -2500 and -3000$\,\mathrm{fs}^2/\mathrm{mm}$, once the gain dispersion is included (defined by \ac{lef}). In sharp contrast to this, \ac{nb} solitons are typically obtained for $k^{''}$ values between 500 and 1200$\,\mathrm{fs}^2/\mathrm{mm}$ -- meaning that homoclons and \ac{nb} solitons are found in different regions of the laser parameter space. In fact, \ac{nb} solitons coexist with a \ac{cw} single-mode operation inside the \ac{cw} linearly-stable region of the parameter space, while homoclons can appear only inside the \ac{cw} linearly-unstable region where turbulence destabilizes the single-mode field.

Homoclons can be classified easily by observing their intermode phase distribution. As is seen from Fig.~\ref{sup_theoryHomo}, the intermode phases form two clusters separated by $\pi$. In the extreme case where the homoclon exhibits only one pulse-like feature per roundtrip (Fig~\ref{sup_theoryHomo}d)), there is a single $\pi$ jump located at the position of the strongest mode.
\ac{nb} solitons, on the other hand, always exhibit two $\pi$ jumps in the intermode phases around the primary mode.
Furthermore, the spectral bandwidth of homoclons is narrower compared to \ac{nb} soliotns, which is why the amplitude variations in their temporal waveform are shallow.

\begin{figure*}[h]
	\centering
	\includegraphics[width = 0.4\textwidth]{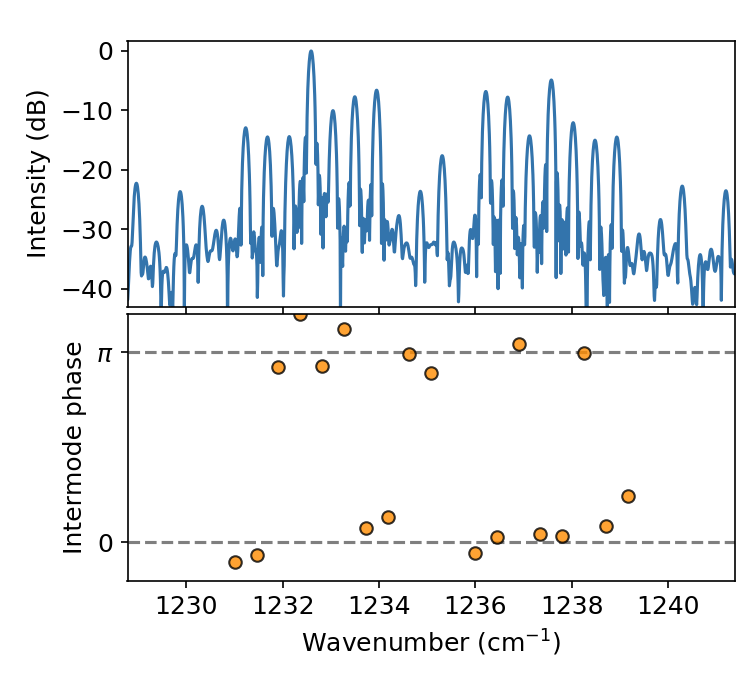}
	\caption{ \textbf{Experimental homoclon state.} The intensity spectrum is shown on top. The corresponding intermode phases are grouped in two clusters separated by $\pi$.  } 
	\label{sup_expHomo}
\end{figure*}

Experimental evidence of a homoclon is shown in Fig.~\ref{sup_expHomo}. The state is obtained for bias currents of $J_{\mathrm{R}}=$1.39$\,\mathrm{kA/cm}^2$ and $J_{\mathrm{WG}}=$0.86$\,\mathrm{kA/cm}^2$, from the same device which emits the \ac{nb} solitons shown in the main manuscript. This indicates that bias tuning of our devices provides a powerful knob that controls the \ac{gvd} over a large range of values, enabled by the separate electrical contacts of the ring and the waveguide coupler.

\section{Coupled mode theory for laser dispersion analysis}
\label{sec:cmt}

The experimental verification of homoclon states (section~\ref{sec:homoclons}) and also the spectral shift of \ac{nb} solitons (section~(\ref{sec:shifted_sol}) provide two persuasive arguments that we are able to tune the \ac{gvd} over a vast range of values purely by changing the ring and/or waveguide current. In this section, we will corroborate this hypothesis with theoretical evidence as well.

Mode interactions between the ring and waveguide are well described by the \ac{cmt}.  Using \ac{cmt}, we can estimate how much the electric field leaks from one resonator (the ring) to the other resonator (the waveguide) when the two are in close proximity to one another.  For devices reported in this work, the ring and waveguide are separated by a narrow air gap ($\sim 0.8 \, \upmu$m) across an interaction length $L_{\text{int}}$ ($\sim 1500 \, \upmu$m).
Bearing in mind that all of the light is generated in the ring, we write mathematical expressions for the electric field in the ring ($E_{\text{R}}$), and the electric field localized in the waveguide ($E_{\text{WG}}$),  after a propagation length $z$~\cite{kazakov2022semiconductor}
\begin{align}
E_{\text{R}}(z)  &= E_0 \left[ \cos{\left(\psi z\right)} + \frac{i\delta}{\psi}\sin{\left(\psi z\right)} \right]\exp{\left(-i\delta z\right)} \\
E_{\text{WG}}(z)  &= -E_0 \frac{i\kappa}{\psi}\sin{\left(\psi z\right)}\exp{\left(i\delta z\right)},
\end{align}
where $E_0$ is the initial electric field amplitude, $\delta$ is the difference of the propagation constants in the ring and the waveguide, $\kappa$ is the difference in propagation constants of the two even/odd super-modes formed in the coupled resonators, and $\psi = \sqrt{\kappa^2 + \delta^2}$. 
$E_{\text{R}}(L_{\text{int}})$ is then used to calculate the \ac{gdd} of the ring as function of frequency $\omega$, which is given by
\begin{align}
\text{GDD}_{\text{R}}(\omega) = \frac{\partial^2\phi_{\text{R}}}{\partial \omega^2},
\end{align}
where $\phi_{\text{R}}$ is phase angle of $E_{\text{R}}(L_\text{int})$.

The propagation constants of the ring and waveguide drift from one another when the two sections are biased at different pumping levels. By measuring the transmission of a weak probe laser through the ring-waveguide system while varying their pumping below threshold and fitting the resultant signal to a transfer matrix model, we extract how the refractive indices (and therefore, propagation constants) change as a function of injection current.  We estimate the change in the refractive index as a function of pumping to be of the order $\sim 0.01\ \text{cm}^{2} \ \text{kA}^{-1}$~\cite{kazakov2022semiconductor}, corresponding to $\delta n \sim 0.001 - 0.010$ for practical pumping levels and typical device geometries.  Furthermore, we use a commercial eigenmode solver to calculate the even and odd super-mode indices for a typical laser cross-section, estimating their difference to be $\sim 0.007$ for an air gap of $0.8$ $\mu$m. 

\begin{figure*}[h]
	\centering
	\includegraphics[width = 0.67\textwidth]{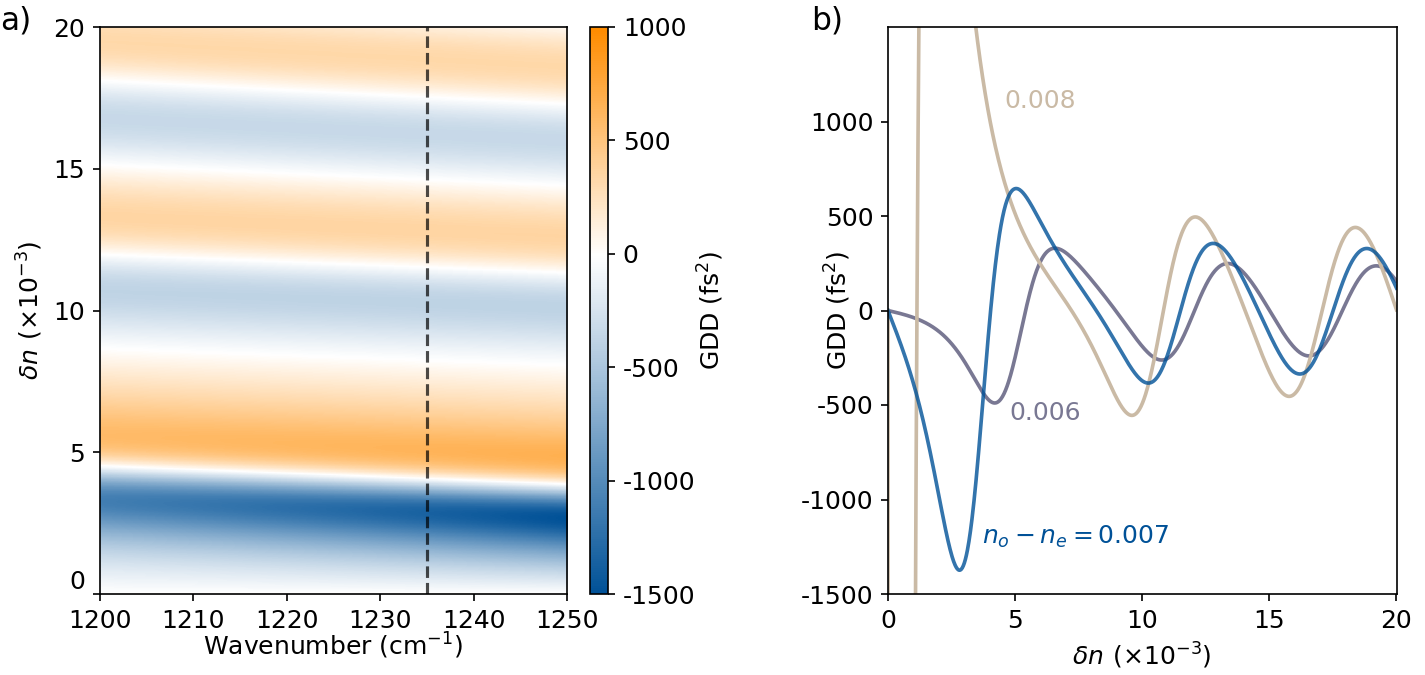}
	\caption{ \textbf{Coupled mode theory analysis of the dispersion.} \textbf{a)} A two-dimensional heat-map of the \ac{gdd} in the ring as a function of frequency $\omega$ and refractive index difference $\delta n$ ($n_{\text{e}} - n_{\text{o}} = 0.007$).  \textbf{b)} A cut of the heat-map at $\omega = 1235 \ \text{cm}^{-1}$ for different values of $n_{\text{e}} - n_{\text{o}}$ (0.006, 0.007, and 0.008).}
	\label{sup_cmt}
\end{figure*}

Fig.~\ref{sup_cmt}(a) shows a calculated two-dimensional heat-map for $\text{GDD}_{\text{R}}$ as a function of both frequency and refractive index difference, spanning from 1200 $\text{cm}^{-1}$ to 1250 $\text{cm}^{-1}$ in frequency, and 0.000 to 0.020 in index difference.  Fig.\ref{sup_cmt}(b) shows a cut of the heat-map at 1235 $\text{cm}^{-1}$ -- the center of the laser gain bandwidth, for different values of super-mode index differences (0.006, 0.007, and 0.008).  Note that regardless of the super-mode index difference, the dispersion changes from normal (negative \ac{gdd}) to anomalous (positive \ac{gdd}) as $\delta n$ is swept.  Therefore, slight pumping differences between the ring and waveguide can have drastic effects on device dispersion.
As a conclusion, our cavity geometry, comprising a coupled ring-waveguide system with separate electrical contacts, provides a powerful knob to control the total mode dispersion -- far superior compared to a single ring or waveguide cavity.

\section{Dark- and bright-pulse NB solitons}
\label{sec:bright}

\begin{figure*}[h]
	\centering
	\includegraphics[width = 0.65\textwidth]{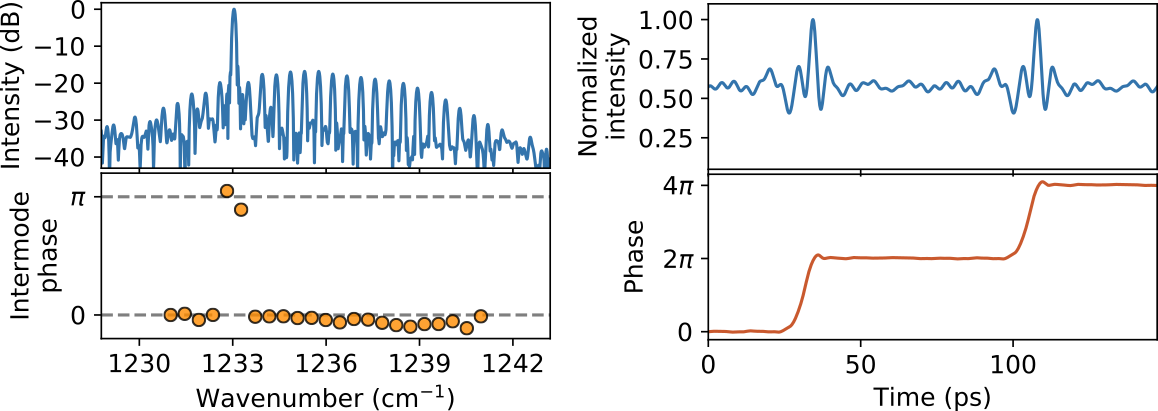}
	\caption{ \textbf{Experimental evidence of a coherent bright pulse.} Detailed experimental characterization of a state from Fig.~4e) in the main manuscript, obtained at bias currents of $J_{\mathrm{R}}=$1.39$\,\mathrm{kA/cm}^2$ and $J_{\mathrm{WG}}=$0.85$\,\mathrm{kA/cm}^2$. }
	\label{sup_justBright}
\end{figure*}

The detailed experimental characterization of the state from Fig.~4e) in the main manuscript can be seen in Fig.~\ref{sup_justBright}. As was explained in the main manuscript, the obtained state looks similar to the dark \ac{nb} soliton state from Fig.~\ref{sup_nb} (Fig.~2 of the main manuscript), with the temporal intensity being the exception as it strikingly shows a formation of a bright coherent pulse. It was argued that this is due to the different intensity of the soliton sidemodes relative to the primary mode, where this ratio is higher in the case of a bright pulse than in the case of the dark pulse. Here we will elaborate more on this.

\begin{figure*}[t]
	\centering
	\includegraphics[width = 1\textwidth]{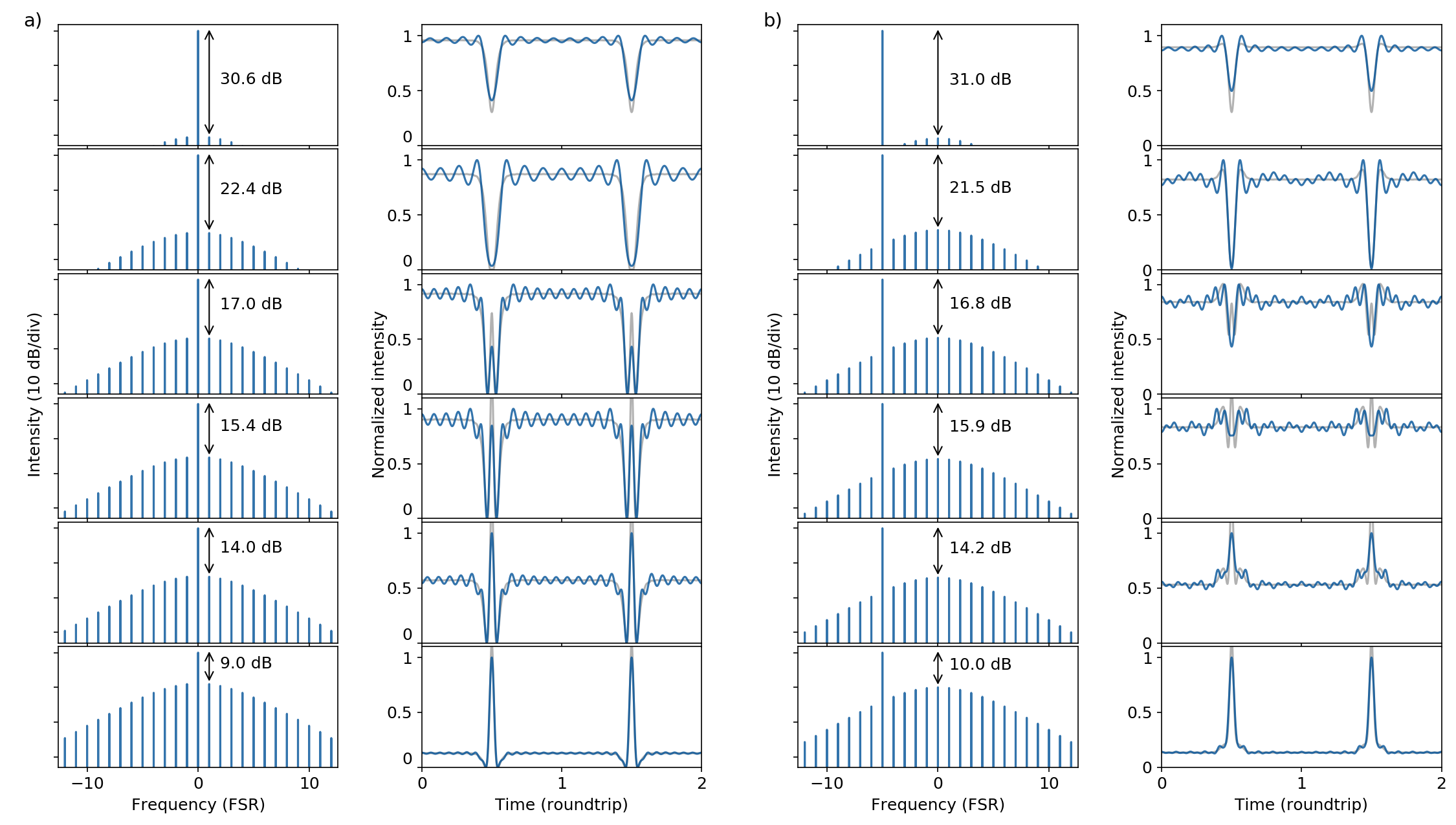}
	\caption{ \textbf{Theoretical analysis of the dependence of the intensity waveform on the evolution of the soliton spectrum.} The primary mode is positioned \textbf{a)} at the center of the soliton spectral envelope, \textbf{b)} away from the center of the soliton spectral envelope. We define the soliton sidemodes using a $\text{sech}^2$ distribution, whose choice is purely arbitrary and only serves to illustrate a trend. The intermode phases are assumed to have $\pi$ jumps around the primary mode and be identical everywhere else. Blue line represents the intensity waveform reconstructed from the modes that are visible in the corresponding displayed spectrum, while the grey line is reconstructed using the weaker modes as well. The blue line exhibits residual oscillations, just like experimentally-recorded intensity waveforms, which is a consequence of the limited number of terms in the Fourier series used for the temporal waveform reconstruction in equation~(\ref{eq:sum}).
		The behavior in both cases is similar, except that for a symmetric spectrum, the intensity reaches zero value at certain point where the dark pulse effectively 'flips over' and becomes bright. }
	\label{sup_contrastEvolution}
\end{figure*}

The theoretical analysis from Fig.~4f) in the main manuscript is displayed in detail in Fig.~\ref{sup_contrastEvolution}. Two cases are investigated, when the primary mode is in the center of the spectral soliton envelope, and shifted from it, which corresponds more to the experimental spectra. Both cases agree with the analysis from the main manuscript. Initial weak soliton sidemodes correspond to a shallow dark pulse in the intensity waveform, whose contrast grows if we increase the sidemode amplitudes relative to the primary mode. At one point (second row in Fig.~\ref{sup_contrastEvolution}), the destructive interference between the primary mode, which defines the CW background, and the sidemodes is complete, resulting in zero intensity at the pulse minima. Further increase of the sidemode amplitudes results in the spectral soliton envelope having a larger intensity compared to the primary mode, at which point the dark pulse would effectively 'flip over'. This is especially evident from the evolution in Fig.~\ref{sup_contrastEvolution}a), where the intensity reaches zero before the pulse 'flips over'. Further increase of the sidemode amplitudes decreases the amplitude contrast of the dark pulse (third row), until a quasi-constant intenisty is obtained (fourth row).
Additionally amplifying the sidemodes above this point results in the potential emission of high-contrast bright pulses.

\begin{acronym}
	
	\acro{mir}[mid-IR]{mid-infrared}
	\acro{qwip}[QWIP]{quantum well infrared photodetector}
	\acro{fsr}[FSR]{free spectral range}
	\acro{lo}[LO]{local oscillator}
	\acro{ftir}[FTIR]{Fourier transform infrared}
	\acro{swifts}[SWIFTS]{Shifted-Wave Interference Fourier Transform
		Spectroscopy}
	\acro{rf}[RF]{radio-frequency}
	\acro{nb}[NB]{Nozaki-Bekki}
	\acro{cw}[CW]{continuous-wave}
	\acro{qcl}[QCL]{quantum cascade laser}
	\acro{fp}[FP]{Fabry-P\'{e}rot}
	\acro{shb}[SHB]{spatial hole burning}
	\acro{cgle}[CGLE]{complex Ginzburg-Landau equation}
	\acro{gvd}[GVD]{group velocity dispersion}
	\acro{lef}[LEF]{linewidth enhancement factor}
	\acro{rt}[R]{ring}
	\acro{wg}[WG]{waveguide}
	\acro{gdd}[GDD]{group delay dispersion}
	\acro{cmt}[CMT]{coupled mode theory}
\end{acronym}

\end{document}